# Priority to Unemployed Immigrants? A Causal Machine Learning Evaluation of Training in Belgium

Bart Cockx[$], Michael Lechner[$$], Joost Bollens[$$$*]

[$] Department of Economics, Ghent University

[$$] Swiss Institute for Empirical Economic Research (SEW), University of St. Gallen

[$$$] Vlaamse Dienst voor Arbeidsbemiddeling en Beroepsopleiding (VDAB)

**Abstract:** Based on administrative data on unemployed in Belgium, we estimate the labour market effects of three training programmes at various aggregation levels using Modified Causal Forests, a causal machine learning estimator. While all programmes have positive effects after the lock-in period, we find substantial heterogeneity in effectiveness across programmes and unemployed. Simulations show that "black-box" reassignment rules that respect capacity constraints on average, increase, respectively decrease, the time spent in employment, respectively unemployment, by more than one month within 30 months of programme start. A shallow policy tree delivers a simple rule that realizes about 85% of this gain.

**Keywords**: Policy evaluation, active labour market policy, causal machine learning, modified causal forest, conditional average treatment effects.

**JEL classification:** C21, J68.

**Addresses for correspondence**:

Bart Cockx, Ghent University, Department of Economics, Sint-Pietersplein 6, 9000 Gent, Belgium, bart.cockx@ugent.be, sites.google.com/site/bartcockxsite.

Michael Lechner, Swiss Institute for Empirical Economic Research (SEW), University of St. Gallen, Varnbüelstr. 14, 9000 St. Gallen Switzerland, Michael.Lechner@unisg.ch, www.michael-lechner.eu, www.researchgate.net/profile/Michael_Lechner.

---

[*] Bart Cockx is also affiliated with IZA Bonn, CESIfo, Munich, IRES, Université catholique de Louvain and ROA, Maastricht University. Michael Lechner is also affiliated with CEPR, London, CESIfo, Munich, IAB, Nuremberg, IZA, Bonn, and RWI, Essen. Michael Lechner gratefully acknowledges financial support from a grant of the NRP 75 of the Swiss Science Foundation (grant SNSF 407540_166999). We thank Michael Knaus, Jeff Smith, Anthony Strittmatter and Gabriel Okasa for carefully reading and commenting on a previous draft of the paper. Previous versions of the paper were presented at research seminars of the LMU University of Munich, the University of Milan, the Competence Centre of Microeconomic Evaluation of the Joint Research Centre of the European Commission, and at the IMT in Lucca, as well as in workshops and conferences in Copenhagen, St Gallen, Engelberg, and Basel. We thank participants, in particular Bas Van der Klaauw, Claudio Lucifora, Regina Riphahn, Claus Schnabel, and Andreas Steinmayr, two anonymous reviewers, and Albrecht Glitz, the editor, for helpful comments and suggestions. We are grateful to the VDAB for providing us access to the data. The VDAB stocks the data and has engaged to make the data available to researchers who wish to replicate the current research. All computations have been performed with the Python MCF package (version 0.2.0 and 0.3.1), which is freely available from PyPI via *pip install*. The usual disclaimer applies.

# 1 Introduction

Unemployment remains an important economic and social concern in Europe, even though in the European Union (EU) the (overall) unemployment rate has steadily decreased from 10.9% in 2013 to 6.8% in 2018 (Eurostat). This is particularly true for some vulnerable groups, such as youth, older workers, and migrants. Policymakers, therefore, have continued interest in getting a better understanding of which labour market policies work for whom. Such understanding helps to improve the counselling process, the design, and the allocation of active labour market policies.

Getting a better understanding of which programme works for whom, and subsequently determining a rule that can improve the allocation of individuals to programmes is challenging. First, estimators of programme effects are typically not designed to flexibly estimate treatment effects at the fine-grained individual level.[1] If they do, they are likely to use some parametric model for the necessary smoothing to avoid the curse of dimensionality (e.g., as in Lechner and Smith, 2007), but this then risks biasing the estimator. Second, it is difficult to derive a general analytical rule that determines which individuals should be assigned (or not) to which programme to maximize a criterion, such as the overall time that these individuals are employed. Recent developments in causal machine learning (CML) propose methods that facilitate the quest for such heterogeneity analysis and assignment rules. In this paper, we use such a CML approach to estimate, at various levels of aggregation, the heterogeneity in the effectiveness of training programmes in Flanders, a region in the north of Belgium. We then use our estimates to simulate the extent to which the public employment service (PES) can enhance the effectiveness of these programmes by changing the assignment of unemployed job seekers to these programmes. We demonstrate that such reassignment can substantially improve allocations, demonstrating the value of using such CML methods for the evaluation of active labour market programmes.

[1] While methods allow for treatment effect heterogeneity in the estimation of average treatment effects, they typically do not aim at estimating average treatment effects conditional on a fine-grained combination of explanatory variables at the individual level, the so-called *conditional* or *individual average treatment effects* (CATE or IATE).

Machine learning methods are traditionally used for prediction (e.g., Hastie, Tibshirani, and Friedman, 2009). More recently, these methods have been modified such that they are useful for causal inference as well (see Athey, 2019, and Athey and Imbens, 2019, for overviews). This literature shows how the counterfactual causal problem (e.g., Imbens and Wooldridge, 2009) can be transformed into a combination of specific prediction problems. For this paper, these methods are of interest because they provide a way to uncover the underlying heterogeneity of the causal effects, a goal for which traditional econometric methods fail to provide a systematic solution. At the same time, we will show that the causal machine learning methods will provide very similar and, hence, reliable estimates of the *average* treatment effects as the traditional methods based on propensity score matching.

In this paper identification of the causal effect relies on the assumption of unconfoundedness. Estimation is based on the Modified Causal Forest (MCF) estimator proposed by Lechner and Mareckova (2022). This builds on the Causal Forest estimators proposed by Wager and Athey (2018). The main innovation is that Lechner and Mareckova (2022) improve on the objective function used to build the trees of the Causal Forest leading to improve performance. In addition, by using weight-based inference methods to estimate the approximate precision of the estimated treatment effects at the various aggregation levels of interest, from the individualized to the (grouped) average treatment effects, estimation and computation are simplified. Nevertheless, these methods are still vulnerable to biases induced by multiple testing. Readers must be aware that there exists currently no satisfactory solution to this problem (see Section 4.4.3).

To the best of our knowledge, this paper is one of the first papers that applies CML methods to analyse treatment effect heterogeneity in the evaluation of active labour market policies.

It appears also to be the first paper to perform such analysis in a multiple treatment context. Knaus, Lechner and Strittmatter (2020) use LASSO-based methods to evaluate the effect heterogeneity of a (single) job search programme in Switzerland using administrative data from 2003. They find substantial effect heterogeneity, but only during the first 6 months after the start of programme participation. Bertrand, Crépon, Marguerie and Premand (2017) apply the Causal Forest method of Wager and Athey (2018) within a randomized controlled trial (RCT) to evaluate the effect heterogeneity of a temporary public works programme in a less developed country. Their analysis reveals important heterogeneities, but again mostly during programme participation.

Unlike the aforementioned papers, this paper considers multiple programmes and explicitly exploits the estimated effect heterogeneity to propose reallocation rules that could increase the programme performance. Two approaches are considered. A first approach aims at finding a rule that reassigns the unemployed to programmes based on the estimated individual programme effects such that it maximizes a chosen objective – here, with equal weight, respectively maximizing and minimizing the time spent in employment and unemployment over a 30-month horizon. One practical disadvantage of this "black-box" approach is that it is not transparent, making caseworkers reluctant to implement such rules (e.g., Lechner and Smith, 2007). In addition, it may conceal that the rule implicitly raises ethical or societal concerns (e.g., Pope and Sydnor, 2011; Whittlestone, Nyrup, Alexandrova, Dihal, and Cave, 2019; Rambachan, Kleinberg, Mullainathan, and Ludwig, 2021). We therefore also consider an approach recently proposed by Zhou, Athey, and Wager (2022) in which optimal policy rules are derived using shallow decision trees with a small number of nodes, resulting in simple rules that are easy to comprehend.

The empirical analysis uses administrative data of the Flemish PES and is based on the population of about 60,000 individuals aged between 21 and 55 who started claiming unem-ployment insurance benefits after an involuntary lay-off between December 2014 and June 2016. To consider labour market outcomes during at least 2.5 years (until September 2019), we evaluate the impact of participating in a subset of training programmes that were entered within the first 9 months of the unemployment spell. We focus the analysis on three training programmes: short-term vocational training (less than 6 months, SVT), longer-term vocational training (between 6 and 10 months, LVT), and orientation training (OT) that aims at helping to determine a clear occupational goal. We could only analyse a subset of the offered training programmes, for various reasons.[2] Consequently, by restricting the set of programmes that the reallocation rules consider, we provide a lower bound to the gains that a reassignment of unemployed to programmes can establish.

The interest in using data from the Flemish PES is threefold. First, the types of training offered to the unemployed are like those offered in most other EU countries, so that findings for Flanders are of general interest. Second, the administrative data is very informative. In addition to rich socio-demographic information, the data contains extensive information on labour market histories (including sickness and past programme participation) of individuals since 1991. This makes the assumption of unconfoundedness, which is used for identification, arguably plausible (see, e.g., Lechner, Miquel and Wunsch 2011, and Biewen, Fitzenberger, Osikominu and Paul, 2014). In fact, in a placebo exercise unconfoundedness could not be rejected (see Section 7.1). Third, the Flemish PES displayed a high willingness to employ CML methods in future programme evaluations and assignments.

The main findings of this paper can be summarized as follows. There is a clear dominance ordering in terms of the average effectiveness of the three programmes considered, both in the short-run (lock-in effects) as well as in the medium run (post-programme effects): SVT performs best, followed by LVT. Although OT also shows positive post-programme effects, the lock-in effects are so large and enduring, that the overall effect in our observation period is negative. After 2.5 years, participation in SVT increases on average the time spent in employment by 2.5 months relative to no participation. For LVT this gain is close to zero, while participation in OT decreases the number of months in employment by 1.4. There is considerable heterogeneity in the aforementioned effects. The effects are especially higher for recent migrants with low proficiency in Dutch, which is the official language in Flanders.

[2] Other programs were not considered because (i) they did not pass a placebo test; (ii) Dutch language training because of lack of comparison observations (common support problem); (iii) they lasted too long (more than 10 months) given the time horizon of 2.5 years; (iv) on-the-job training, because participants were selected after hiring; or (v) they were too small and too heterogeneous to be aggregated in a meaningful group.

Finally, we study to which extent the Flemish PES could improve the effectiveness of the training component of their ALMP by simulations that change the allocation of programme participants according to the individualized effectiveness of these programmes, following the two aforementioned approaches. Based on the black-box approach, we find that a reallocation that keeps programme capacity fixed could for the group affected by this reassignment on average increase, respectively decrease, the time spent in employment, respectively unemployment, by about one month and one week within 30 months. Shallow policy trees of depths 2 to 4 define policy rules that can achieve about 85% of this gain. These rules reveal that most gains are achieved by reallocating workers born abroad and with little employment experience in Belgium (i.e., typical profiles of recent immigrants) to short vocational training programmes. By contrast, it assigns older workers displaced from specific sectors and with little recent employment experience to long vocational programmes, consistent with the idea that such workers need intensive retraining for their professional reorientation. Finally, nobody would be reassigned to orientation training, because this programme is for most unemployed less effective than remaining unenrolled.

The rest of the paper is structured as follows. In Sections 2 and 3 we describe the institutional setting and the data that are used in the analysis. Section 4 discusses the econometric methods. Section 5 presents the results with a focus on the analysis of effect heterogeneity. Section 6 simulates several alternative assignment rules. It is followed by some robustness analysis, including the placebo analysis, and ends with concluding remarks. An appendix provides additional details on the data, the estimator used, the results, and the sensitivity analysis.

## 2 The institutional setting

Belgium is a federal state in which many competencies have been decentralized. Generally, location-based matters, such as employment policies, are decentralized to the three regions (Flanders, Wallonia, and Brussels). The Regional Public Employment Services (PES) oversee job search assistance, intermediation services and the provision of active labour market policies (ALMP) for the unemployed. The rules and payment of unemployment insurance (UI) are determined at the federal level. We first explain the entitlement rules to UI at the federal level and then explain how the assignment to ALMP is organised in the Flemish PES, the focus of our analysis in this paper.

Unemployed workers are entitled to non-means-tested unemployment benefits (UB) in two cases. First, graduates from high school or higher education who are younger than 25 can start claiming benefits after a waiting period of one year. Second, workers with recent experience of at least one year in wage and salary employment qualify for UB after an involuntary lay-off.[3] We focus in this paper on the latter scheme. Because of our finding that programme participation is most effective for recent immigrants, it is important to note that recent immigrants could qualify for unemployment benefits in Belgium based on proof of equally long periods of wage and salary employment abroad provided that at least one day was worked in Belgium.[4]

[3] Workers older than 35 must have worked more than one year to qualify for UB.

Unlike in other countries, the benefits are paid out without a time limit. The UB is related to prior earnings but bracketed by a cap and a floor. The replacement rate is initially 65% (with a maximum of €1,736/month), but declines with unemployment duration. It drops to 60% after 3 months, and then further after one year, depending on the status within the household and prior work experience. After 4 years all UB attain the minimum which depends on the household status: €1,316/month for heads of household, €1,078/month for singles, and €561 for dependents, before taxes.[5]

In Flanders, the assignment to ALMP occurs as follows. Soon after registration as a job seeker, the unemployed is invited to an intake meeting or phone call with a counsellor. At this intake meeting information is provided concerning rights and duties, and, based on the available information in the register files (socio-demographic information and labour market history described in Section 3) and the information provided by the job seeker, the counsellor makes a subjective assessment whether the individual can autonomously find employment without any help from the PES, or not.[6] Job seekers assessed to need PES services are then subsequently invited to a second meeting with a counsellor to determine which services are best suited to the needs of the job seeker.

For determining the needs of job seekers in these two meetings, counsellors are trained to systematically take the following criteria into account:

---

[4] See RVA/ONEM in Dutch, French or German (accessed on 2020-04-27). This rule changed for individuals claiming UB from November 2016 onwards, but this is not relevant for our analysis as we retain only individuals flowing into unemployment until June 2016 (see Section 3.1).

[5] Amounts for July 2019 (https://www.rva.be/nl/documentatie/infoblad/t67).

[6] Before December 2014 this assessment was not systematically made for all job seekers. Therefore, we consider only inflows into unemployment from this date onwards (see Section 3.1).

(i) Is there a clear and realistic job target?

(ii) Are there any obstacles preventing a transition to employment, related to health, mobility barriers, inadequate search behaviour or work attitudes (such as getting up in time, or accepting authority), insufficient knowledge of the labour market, and the like?

(iii) Is the job seeker sufficiently proficient in Dutch, the local language in Flanders?

(iv) Does the job seeker have the required competencies for exercising the chosen profession?

Depending on the answers to the aforementioned questions, the counsellor assigns the job seeker to a different programme. Job seekers without a clear or realistic job target (criterion (i)) are recommended to follow *orientation training* (OT), those facing some obstacle (criterion (ii)) are proposed to follow *intensive counselling* (IC), and those lacking proficiency in Dutch (criterion (iii)) are referred to a *language training* (DLT), or they may enter another programme with the support of a language coach. Unemployed lacking the required competencies for a chosen profession (criterion (iv)) can be oriented to *vocational training* (VT) for certain occupations, or *on-the-job training* (OJT). VT and OJT are available for a wide range of professions, and for some of these programmes, admission requires a degree from a particular educational track or experience in a similar profession. Finally, counsellors typically only assign workers to programmes if they are sufficiently motivated.

Allegedly,[7] most counsellors typically assign only motivated job seekers to an ALMP and do not impose participation, as this is not required by the rules. On the other hand, participation in such programmes is accepted as proof of compliance with the job search requirements for UI claimants. This incentivizes programme participation. Moreover, once agreed to participate in a programme, a job seeker is obliged to complete it. Unemployed workers who end participation prematurely risk a benefit sanction.

[7] We are not aware of any research that can substantiate this claim.

Information about the delay between the assignment and the actual programme start is not available. However, for short and long vocational training the median delay is shorter than 32 days, as this corresponds to the median duration between the assignment and the start of each of these programmes.[8] Except for OT, which is often followed by participation in another programme, repeated assignments to different programmes within the same unemployment spell are not commonplace.

Finally, we underline that the Flemish institutional setting for the provision of ALMP is not very different from that in many other developed countries. Moreover, the level of ALMP expenditures is also very comparable (see OECD.stat). In 2016, the Belgian expenditures on labour market services and training amounted to 0.35% of GDP. This is somewhat higher than the OECD average of 0.25% and higher than the 0.30% for the neighbouring country, the Netherlands. The other neighbours, France, and Germany spend more, respectively 0.54% and 0.55% of GDP.

# 3 Data

## 3.1 Population of interest

The data for the analysis is drawn from administrative files on all individuals who registered between January 1991 and February 2019 as unemployed job seekers at the Flemish PES. These files contain rich socio-demographic information, as well as individual employment and job search histories since 1991. From this database, we select 148,942 individuals who started claiming UI after an involuntary lay-off between December 2014 and June 2016. We do not retain individuals who enter unemployment before December 2014, because before this date the selection into training programmes was different (see Section 2) and not after June 2016 to allow for follow-up of participants over a sufficiently long period. Since we retained programmes that commenced up to 9 months after the start of the unemployment spell and since the observation period ends in September 2019, labour market outcomes can be observed for up to 30 months after the programme start.

[8] In principle, the programme start should be defined from the assignment to a programme because the individual can from that moment onwards change job search behaviour in anticipation of the training. However, since the median delay is less than a month, and the data are grouped in monthly intervals, this measurement issue cannot result in any significant bias in the estimated treatment effects. We have no information about this delay for OT but assume that it is of a similar magnitude.

We exclude school-leavers claiming UB (see Section 3) as well as individuals younger than 21. We also exclude workers with disabilities and those older than 55 at the start of the unemployment spell, because these groups need not be fully available for the labour market or may benefit from alternative policies. We also dropped individuals not living in Flanders and those who died during the period of analysis. 73,582 individuals are retained after imposing these selection criteria.

## 3.2 Programmes

Table 3.1 reports how this population is divided up into four subgroups: (1) 56,324 individuals who did not participate in any ALMP within the first 9 months of unemployment, i.e. the not yet treated group (Sianesi, 2004); (2) 3,640 individuals who started within the first 9 months an ALMP retained in the analysis; (3) 13,618 individuals who entered within the first 9 months an ALMP and who are not retained for the evaluation, because (i) the placebo tests suggested concerns about selection bias for the evaluation of intensive counselling,[9] (ii) only very few foreigners without any knowledge of Dutch participated in other programmes than language training, so that too few comparison observations were available in these cases,[10] (iii)

[9] This placebo test is not reported. It was implemented in the same way as explained in Section 7.1 for the included data.

[10] Note that this is only an issue for foreigners with no knowledge of Dutch, i.e. for whom the level of proficiency is zero. Foreigners with proficiency level one, or above, do participate in the other programmes, such that our finding that

on-the-job training is not comparable to the other ALMP, as it is assigned to individuals who have already found a job and (iv) other small ALMP could not be aggregated in meaningful groups sufficiently large for an empirical analysis, or they belonged to a category that lasted too long on average (10 months or more) for an evaluation of the medium run impacts within the 30 months observation period.[11]

*Table 3.1: Importance of ALMP by type for entry cohorts in unemployment between December 2014 and June 2016*

| **Type of assistance and ALMP** | Average programme duration (months) | Number of individuals | Fraction |
|---|---|---|---|
| *No ALMP participation within the first 9 months (NOP)*[1] | - | 56,324 | 76.5% |
| *ALMP within the first 9 months retained in the main analysis* | | 3,640 | 4.9% |
| 1. Short (< 6 months) vocational training (SVT)[2] | 3.83 | 1,305 | 1.8% |
| 2. Long (< 10 months) vocational training (LVT)[2] | 7.18 | 1,220 | 1.7% |
| 3. Orientation training (OT) | 1.05 | 1,115 | 1.5% |
| *ALMP within the first 9 months excluded from the analysis* | | 13,618 | 18.5% |
| 1.Intensive counselling (IC) | - [4] | 3,695 | 5.0% |
| 2. Dutch language training (DLT) | 2.56 | 991 | 1.3% |
| 3. On-the-job training (OJT) | - | 2,045 | 2.8% |
| 4. Other ALMP, including *very long* VT[2,3] | - | 6,887 | 9.4% |
| Total | | 73,582 | 100.0% |

Note: Retained individuals are aged between 21 and 55 years at registration and started claiming UB after lay-off in the period December 2014-June 2016. The following individuals were excluded: (i) those not living in Flanders; (ii) those with some disability; (iii) those who died during the period of analysis.
[1] This group may enter programmes beyond the first 9 months. About 25% enter an ALMP before month 31.
[2] No information on planned duration is available. The duration of vocational training (VT) is determined as the average realized duration in the corresponding sector. Since the number of VT in some sectors was too small, some of them were aggregated. Eventually, 31 sectors are distinguished, all containing at least 19 individuals.
[3] This group contains various types of small, heterogeneous ALMPs as well as 1,492 individuals who participated in vocational training lasting 10 months or more.
[4] The administrative files only record the administrative end of the contract as determined by the service provider to which the IC is contracted out. The duration of the service provision is not known but must be less than the contract duration which lasted 8.82 months on average.

Programme participants are classified according to the first programme they participate in. The programmes considered in the analysis are the following. First, following Lechner, programmes are particularly effective for recent immigrants is relevant. In Table A.1 in the Online Appendix, it can be verified that the fraction born abroad ranges between 35% for those with no ALMP participation within the first 9 months (NOP) and 29% for those participating in orientation training (OT).

[11] Measuring outcomes of such programmes in the short- or medium-run would not make much sense as they would merely capture the lock-in effects.

Miquel and Wunsch (2011), we distinguish between *short*- (less than 6 months, 3.8 months on average) and *long* (more than 6 and less than 10 months, 7.8 months on average) *vocational training* programmes (SVT and LVT). Very long vocational training programmes, lasting 10 months or more, were reclassified into *other ALMPs,* and subsequently dropped from the analysis to ensure a sufficiently long follow-up. Since no information on planned duration is available, the duration of vocational training (VT) is determined based on the average realized duration in the corresponding sector. We stress that these programmes do not only differ by duration, but also by content. SVT and LVT both aggregate training for different occupations in different sectors. LVT is therefore not just more of SVT.

Third, *orientation training* (OT) aims at helping the unemployed to determine a clear occupational goal. This programme is relatively short. It lasts on average only about one month. However, within three months after the end, 45 per cent of the participants enter another ALMP, presumably to support the orientation that they have chosen.

### 3.3 Confounding and outcome variables

The dataset contains 45 ordered and 9 categorical conditioning variables (with 3 to 44 unordered categories). All time-varying variables are measured at the start of the unemployment spell. Taking into account that usually categorical variables would be transformed into dummy variables in a regression-type setting,[12] this would correspond with 175 and many more if one aims at avoiding parametric restrictions by the inclusion of interaction and higher-order terms that the MCF will automatically account for.

These conditioning variables provide information about personal socio-demographic characteristics, labour market history, including sickness, within the preceding 2, 5 and 10 years, the ALMP participation history during previous unemployment spells, information about the job seeker's job preferences and the corresponding occupational experience, the calendar month in which unemployment was entered (19 indicators) and the day at which the ALMP started (or was predicted to start in case of no participation)[13] in the unemployment spell (maximum 274 days).

[12] This recoding is not needed for the MCF as it treats categorical variables directly, like Random Forests (as in Chou 1991; Hastie et al. 2009/2013, p. 310).

For a correct interpretation of our findings reported below, the reader should be aware of the following data limitations: (i) labour market history is recorded only from the first entry into unemployment, which means that those experiencing a first unemployment spell will have no reported work experience (14%); (ii) as mentioned in Section 2, foreign work experience counts to qualify for UB but is not registered in the data. These limitations imply that for a sizeable fraction of the selected sample (about 30%) less than 12 months of employment is registered in the last two years, which is in most cases not sufficient to qualify for UB.[14] Because we know that all sampled individuals are entitled to UI, this reflects the aforementioned data limitations. These data limitations are further documented in Panel B of Table 3.2 below.

However, these data limitations need not necessarily be a source of bias. First, by combining the information on age and educational attainment, the data implicitly can proxy the employment history of natives who experience a first unemployment spell, because this group has most likely been uninterruptedly employed since leaving education. Second, the missing information about the employment history of recent immigrants is not necessarily problematic, because (i) foreign employment is less valuable than local labour market experience and (ii) the combination of education, age and other socio-demographic information might proxy the missing information quite well. In the placebo analysis reported in Section 7.1, we find no evidence of specific violations for the non-native population of the conditional independence assumption on which the estimator relies (Section 4.2).

[13] More on this in Section 4.4.

[14] The minimum qualifying period for UI is 12 months of employment in the last 21 months, but this is not strict, as earlier employment experience can compensate for the lack of recent experience to the extent that this earlier experience is longer than 12 months.

Table 3.2 reports summary statistics for a selected set of conditioning variables (panel A), variables documenting the aforementioned data limitation regarding the labour market history (panel B), and outcomes (panel C): the sample means by programme status and the standardized differences (in %) for each programme status (SVT, LVT, OT) relative to the NOP group that did not participate in any ALMP within the first 9 months of the unemployment spell. A full description of all explanatory variables and the corresponding statistics can be found in Online Appendix A.1. The standardized differences are often larger than 20%, a number that Rosenbaum and Rubin (1985) consider 'large'. This signals that the conditioning variables of participants in ALMP are very unbalanced relative to the NOP group and that controlling for selection on observed variables is crucial in this setting.

We observe that there are many more men than women participating in *vocational training*, especially in SVT. Participants in vocational training are on average somewhat more proficient in Dutch, especially those in LVT. Participants in SVT have on average comparable unemployment experience as non-participants, while those in LVT have been on average less unemployed both in the last 2 and 10 years. Participants in SVT are on average less educated than non-participants, while those in LVT are on average more educated.

*Table 3.2: Means and standardized differences for selected variables*

| Variable | No ALMP participation (NOP) | Short vocational training (SVT) | | Long vocational training (LVT) | | Orientation training (OT) | |
|---|---|---|---|---|---|---|---|
| *A. Conditioning variables* | Sample mean (*standardized difference*100 relative to NOP*)[1] | | | | | | |
| Woman | 0.49 | 0.31 | *(36)* | 0.40 | *(16)* | 0.46 | *(3)* |
| Age (in years) | 35 | 34 | *(12)* | 34 | *(12)* | 34 | *(16)* |
| Proficiency in Dutch (0-3)[2] | 2.4 | 2.5 | *(8)* | 2.7 | *(40)* | 2.6 | *(27)* |
| Months unemployed in the last 10 years | 18 | 19 | *(5)* | 16 | *(11)* | 17 | *(4)* |
| Months unemployed in last 2 years | 3.9 | 3.8 | *(2)* | 3.0 | *(18)* | 3.2 | *(14)* |
| Education level (1 to 13) | 7.2 | 5.9 | *(38)* | 7.9 | *(22)* | 7.2 | *(1)* |
| *B. Availability of labour market history data (LMHD)*[3] | | | | | | | |
| Fraction of people in first unemployment spell (no LMHD) | 0.14 | 0.07 | *(23)* | 0.05 | *(29)* | 0.06 | *(25)* |
| Fraction of people with at least 2 years of LMHD | 0.80 | 0.89 | *(24)* | 0.88 | *(24)* | 0.88 | *(23)* |
| Fraction of people with at least 10 years of LMHD | 0.41 | 0.44 | *(6)* | 0.46 | *(9)* | 0.46 | *(9)* |
| *C. Outcomes* | | | | | | | |
| # of months employed 10 months after starting ALMP[4] | 4.0 | 3.9 | *(2)* | 2.8 | *(33)* | 2.4 | *(45)* |
| # of months employed 20 months after starting ALMP[4] | 9.8 | 11 | *(16)* | 9.4 | *(5)* | 7.8 | *(27)* |
| # of months employed 30 months after starting ALMP[4] | 16 | 18 | *(24)* | 17 | *(11)* | 15 | *(12)* |
| Number of observations | 56324 | 1305 | | 1220 | | 1115 | |

Notes: [1] The standardized difference is defined as $|\bar{x}^j - \bar{x}^{NOP}|/\sqrt{[Var(x^j) + Var(x^{NOP})]/2} * 100$, where $\bar{x}^j$ and $Var(x^j)$ are the sample mean and variance of the variable $x^j$ for $j \in \{SVT, LVT, OT\}$.
[2] Proficiency in Dutch = 0 if no knowledge; = 1 if limited; = 2 if good; =3 if very good.
[3] This information is implicitly conditioned upon, as it can be obtained by (a combination of) values of conditioning variables (see footnote 6 of Table A.1 in the Online Appendix).
[4] For non-participants in ALMP (NOP) the date at which the ALMP starts (is predicted) (See Section 4.4).

Like the trainees in LVT, participants in *orientation training* are on average more proficient in Dutch than nonparticipants. Moreover, like LVT, they have been on average much less unemployed in the last two years. Their level of education is on average like that of nonparticipants. This profile seems to match the profile of a medium-skilled worker who has been employed in a routine job and has been displaced in the gradual tendency of more polarization of the labour market (see e.g., Autor et al. 2003; Goos et al. 2009). These workers typically require re-orientation to another occupation, because they typically have skills that are no longer in demand.

We further analyse the direction of the selection bias between the different programmes in Online Appendix A.2. From this analysis it can be concluded that participants in SVT are on average less qualified for the labour market than those in LVT, while participants in LVT and OT have on average a similar profile. We will refer to this evidence when discussing the relative average effectiveness of these programmes in Section 5.1.

Panel B documents further the availability of labour market history data. About 14% of the sample consists of individuals who experience a first unemployment spell, and for whom information about labour market history is lacking. For the participants in training, this fraction is much lower. It varies between 5% and 7%, respectively for participants in LVT and SVT. For about 80% (41%) of the sample, the labour market history is known for at least two (ten) years, and for participants in training programmes, these fractions are close to 90% (45%). Given that the existing evaluation literature highlights that controlling for *recent* labour market history matters most, these statistics further support that the aforementioned data limitations are not too severe.

Panel C of Table 3.2 reports the summary statistics for three main outcome variables, namely the cumulative number of months that a worker is employed in the 10, 20 and 30 months after the start of the ALMP. In the empirical part, the last one and additional outcome variables will be considered. Since participation in ALMP can last up to 10 months on average (for long VT), the effects over the first 10 months measure *lock-in* effects for some programmes, while after 30 months the post-program effect, if present, adds to this lock-in effect.

It can be deduced from panel B of Table 3.2 that the outcomes vary substantially by programme status. However, because of the important variability of the conditioning variables (panel A) these descriptive statistics are not necessarily informative about *causal* average programme effects due to possible selection biases. How to draw inference about the effects of these programmes is discussed in the next section.

# 4 Econometrics

## 4.1 The causal modelling framework and the parameters of interest

We use Rubin's (1974) potential outcome language to describe a multiple treatment model under unconfoundedness, or conditional independence (Imbens, 2000, Lechner, 2001).

Let $D$ denote the treatment, which is non-participation or participation in one of the three programmes in our case. Thus, it takes on four different integer values from *0* to *3*. The (potential) outcome of interest that realises under treatment $d$ is denoted by $Y^d$. For each individual, we observe only the particular potential outcome related to the treatment status that the individual has chosen, $y_i = \sum_{d=0}^{3} \underline{1}(d_i = d) y_i^d$ (where $\underline{1}(\cdot)$ denotes the indicator function, which is one if its argument is true and zero otherwise).[15] There are two groups of variables to condition on, $\tilde{X}$ and $Z$. $\tilde{X}$ contains those covariates that are needed to correct for selection bias (confounders), while $Z$ contains variables that define (groups of) population members for which an average causal effect estimate is desired. For identification, $\tilde{X}$ and $Z$ may be discrete, continuous, or both, but for estimation, we will consider discrete $Z$ only. They may overlap in any way. In line with the machine learning literature, we call them 'features' from now on. Denote the union of the two groups of variables by $X$, $X = \{\tilde{X}, Z\}$, $\dim(X) = p$.

Below, we investigate the following average causal effects:

$$IATE(m,l;x) = E(Y^m - Y^l \mid X = x) \ ,$$

$$GATE(m,l;z) = E(Y^m - Y^l \mid Z = z) = \int IATE(m,l;x) f_{X|Z=z}(x) dx \, ,$$

$$ATE(m,l) = E(Y^m - Y^l) = \int IATE(m,l;x) f_X(x) dx \, .$$

The **I**ndividualized **A**verage **T**reatment **E**ffects (IATEs), $IATE(m,l;x)$, measure the mean impact of treatment $m$ compared to treatment $l$ for units with features $x$. The IATEs represent the causal parameters at the finest aggregation level of the features available. On the other extreme, the **A**verage **T**reatment **E**ffects (ATEs) represent the population averages. The

[15] If not obvious otherwise, capital letters denote random variables, and small letters their values. Small letters subscripted by '*i*' denote the value of the respective variable for individual '*i*'.

ATE is the classical parameter investigated in many econometric causal studies. The **G**roup **A**verage **T**reatment **E**ffect (GATE) parameters are in between those two extreme aggregation levels. The analyst preselects the variables *Z* before estimation according to her policy interest. The IATEs and the GATEs are special cases of the so-called **C**onditional **A**verage **T**reatment **E**ffects (CATEs).[16]

## 4.2 Identification

The classical set of unconfoundedness assumptions consists of the following parts (see Imbens, 2000, Lechner 2001):

$$\{Y^0, Y^1, Y^2, Y^3\} \coprod D | X = x, \qquad \forall x \in \chi ; \qquad (CIA)$$

$$0 < P(D = d | X = x) = p_d(x), \qquad \forall x \in \chi, \forall d \in \{0, \ldots, 3\} ; \qquad (CS)$$

$$Y_i^{(D_1, \ldots, D_i, \ldots D_P)} = Y_i^{D_i} \qquad (SUTVA)$$

where $Y_i^{(D_1, \ldots, D_i, \ldots D_P)}$ denotes the potential outcome of individual $i \in \{1, \ldots i, \ldots P\}$ in a population of size $P$ if individuals in this population receive treatments $(D_1, \ldots, D_i, \ldots D_P)$.

The conditional independence assumption (CIA) implies that there are no features other than *X* that jointly influence treatment and potential outcomes (for the values of *X* that are in the support of interest, $\chi$ ). The common support (CS) assumption stipulates that for each value in $\chi$, there must be the possibility to observe all treatments. The stable-unit-treatment-value assumption (SUTVA) implies that the observed value of the outcome does not depend on the treatment allocation of the other population members (ruling out spillover effects). Usually, to have an interesting interpretation of the effects, it is required that *X* is not influenced by the treatment (exogeneity). If this set of assumptions holds, then all IATEs are identified:

$$\begin{aligned} IATE(m,l;x) &= E(Y^m - Y^l \mid X = x) \\ &= E(Y^m - Y^l \mid X = x) \\ &= E(Y^m \mid X = x, D = m) - E(Y^l \mid X = x, D = l) \\ &= E(Y \mid X = x, D = m) - E(Y \mid X = x, D = l) \\ &= IATE(m,l;x); \qquad \forall x \in \chi, \forall m \neq l \in \{0,...,3\}. \end{aligned}$$

[16] Similarly, we can also define these parameters for different treatment groups to obtain, for example, average and group average treatment effects for the treated. Although for the sake of simplicity these parameters are not formally discussed, they are also identified by the assumptions below (see Lechner, 2018).

Since the distributions used for aggregation, $f_{X|Z=z}(x)$ and $f_X(x)$, relate to observable variables (*X*, *Z*, *D*) only, they are identified as well (under standard regularity conditions). This in turn implies that the GATE and ATE parameters are identified.

It is of course important that these conditions are plausible in our study. Let us consider them in turn. In Section 3 we already argued that the availability of a wide range of socio-demographic information and rich information about the labour market history of individuals enhances the plausibility of the CIA. These are essentially the variables identified by other evaluation studies as the most important confounders (e.g., Heckman et al., 1998; Lechner and Wunsch, 2013). These are also the variables available to the caseworker during the interview and thus should be the ones she is mainly basing her decision on. Advantages of our study compared to the training programme evaluation literature are the availability of sickness absence records as well as the unemployment rate in the district of residence. Probably the biggest disadvantage is the lack of earnings histories. However, this may not be so important as earnings are not an outcome variable (and thus earnings records are not needed for the role of pre-treatment outcomes) as well as because proxies for earnings are available, such as education, nationality, the sectors of the previous jobs, the duration of the elapsed unemployment duration at the (potential) programme start as well as the duration of the preceding employment spell (with the qualifications mentioned in Section 3.2), and the preferred occupation of the job seeker. Overall, we conclude that the CIA may be plausible. However, as a safeguard against possible violations, we report in Section 7.1 a placebo study (that does not indicate any violations).

SUTVA is plausibly fulfilled as all programmes considered are rather small compared to the labour force. Dutch language training (DLT) was eliminated from the estimation (Section 3.2), because natives are not eligible for it, leading therefore to a problem of non-overlapping support. By contrast, each unemployed worker is in principle eligible for any of the three training programmes retained in the analysis. Furthermore, we did not detect important common support problems with the retained programmes in the data. We explain in the Online Appendix D in more detail how we impose common support. We there show that we had to drop 19% of the observations, but that this did not have major implications for the estimation results apart from reducing the variability of the IATEs. Finally, the exogeneity of confounding and heterogeneity variables is ensured by measuring all time varying variables at the beginning of the unemployment spell. At that moment, the individual did not know if and when she will enter a training programme.

## 4.3 Estimation

In the recently developed causal machine learning literature (see Athey 2019, and Athey and Imbens, 2019, for overviews) the prediction power of the machine and statistical learning literature (for an overview see, e.g., Hastie, Tibshirani, and Friedman, 2009) is combined with the micro-econometric literature on defining and identifying causal effects (e.g., Imbens and Wooldridge, 2009). Recently, the causal machine learning literature has seen a surge of proposed methods, in particular in epidemiology and econometrics. Knaus, Lechner, and Strittmatter (2021) compare many of those methods systematically to their set-up as well as their performance in a simulation exercise. One conclusion from their paper is that random forest-based estimation approaches outperform alternative estimators. In this paper, we use the Modified Causal Forest (MCF) estimator that was proposed by Lechner and Mareckova (2022).[17] In an Empirical Monte Carlo analysis Lechner and Mareckova (2022) demonstrates that, in a context in which identification relies on the CIA, this estimator performs at least as well as the Causal Forest of Wager and Athey (2018). The MCF estimator also proposes a simple way of performing unified inference at all aggregation levels (IATE, GATE and ATE) and can be easily applied to a multiple, discrete treatment framework as in the empirical application of this research.

The starting point of the causal forest literature is the Causal Tree introduced in a paper by Athey and Imbens (2016). In a Causal Tree, the sample is split sequentially into smaller and smaller strata, in which the values of $X$ become increasingly homogenous, to mitigate selection effects and to uncover effect heterogeneity. Once the splitting is terminated based on some stopping criterion, the treatment effect is computed within each stratum (called a 'leaf') by computing the difference between the mean outcomes of treated and controls (possibly weighted by the conditional-on-$X$ probabilities of being a treated or control observation). However, the literature on regression trees acknowledges that the final leaves may be rather unstable because of the sequential nature of the splits (if the first split is different, the full tree will likely lead to different final strata). A solution to this problem is the so-called Random Forest. The key idea is to induce some randomness into the tree-building process, build many trees, and then average the predictions of the many trees. The induced randomness is generated by using randomly generated subsamples (or bootstrap samples) and by considering for each splitting decision only a random selection of the covariates. Wager and Athey (2018) use this idea to propose Causal Forests, which are based on a collection of Causal Trees with small final leaves.[18] Lechner and Mareckova (2022) develops these ideas further by improving on the splitting rule for the individual trees, and by providing methods to estimate heterogeneous effects for a limited number of discrete policy variables (**G**roup **A**verage **T**reatment **E**ffects, GATE) at low computational costs, in addition to the highly disaggregated effects, the literature focussed on so far (**I**ndividualized **A**verage **T**reatment **E**ffects, IATE). Furthermore, his paper suggests a way of performing inference for all aggregation levels. Finally, the approach applies to a multiple, discrete treatment framework. The main advantage to more conventional estimators for multiple treatments is among others the systematic approach to heterogeneity estimation as well as no need to use specific, parametric functional forms to parametrise any outcome or propensity score model. Compared to other causal machine learning estimators, an important advantage of the MCF is that it provides a framework to estimate all causal parameters of interest in one unified estimation.

[17] The specific version is using the penalty function.

[18] Athey, Tibshirani, and Wager (2019) generalize this idea to many different econometric estimation problems.

The interested reader is referred to Lechner and Mareckova (2022) for all further technical details of the estimator and to Bodory, Busshoff, and Lechner (2022) for more details on implementational issues of the Python free package (*mcf*)

## 4.4 Practical implementation

### 4.4.1 Differential programme starts

Because individuals could be assigned to an ALMP at any point of time in their unemployment spell (although usually they are assigned in the beginning), we face a dynamic assignment problem. The problem is that because by construction the probability of enrolment into programmes increases over time, the treatment status depends on unemployment duration, which is (a function) of the outcomes of interest, and this confounds any static analysis which does not take this dynamic feature into account. In this section, we review the literature on this issue and propose to deal with this issue by one of the solutions that have been proposed to deal with this issue within a static framework. However, we also recognize that this solution is imperfect and that it might bias the estimator of the treatment effect downwards, as has been recently shown by van den Berg and Vikström (2022). The latter authors propose an alternative approach to correct this bias. However, as it remains to be shown how this approach can be implemented within a CML framework, it is beyond the scope of this paper to implement it.

A first remark is that dynamic assignment requires an assumption of no anticipation in addition to the CIA. No anticipation means that individuals do not alter their behaviour in response to a future assignment to the ALMP. Since in the period of analysis the training capacity tended to exceed demand, the time between assignment and the actual programme start is short, such that the bias induced by any failure of this assumption is likely to be small.

To transform a dynamic programme assignment into a static one, non-participants are defined to be the population that did not participate in the programme within a certain period, such as the first nine months in this paper. Fredriksson and Johansson (2008) explain that such a definition biases the estimation of the effects downwards, as nonparticipants are less likely to have entered a programme, because they may have already found a job. To avoid this bias, they propose to define the comparison group as those that have not yet been treated. Based on these insights, two strands have developed in the literature. A first strand aims at identifying the effects of those who did not *yet* receive treatment (e.g., Sianesi 2004, 2008 and Biewen, Fitzenberger, Osikominu and Paul 2014). A disadvantage of this approach is that it redefines the effect and makes it dependent on the fraction of nonparticipants that participate (shortly) after this period.[19] Another strand of the literature, therefore, aims at identifying the effect relative to never receiving the treatment. This is essentially done by right censoring nonparticipants who subsequently enter the programme. Fredriksson and Johanson (2008) assume *independent* right censoring, while Crépon, Ferracci, Jolivet and van den Berg (2009) and Vikström (2017) gen-eralize this by allowing for *selective* right censoring. Van den Berg and Vikström (2022) consider long-run post-treatment effects, such as those of vocational programmes on earnings.

[19] In our empirical application, about 25% of the nonparticipants enter an ALMP between 10 and 30 months after the beginning of the unemployment spell.

Identifying the effects relative to never receiving the treatment with CML methods is beyond the scope of this paper. Here, we follow the first strand in this literature. We essentially follow the approach proposed by Lechner, Miquel and Wunsch (2011), which adapts the one suggested by Lechner (1999, 2002), to accommodate the critique of Fredriksson and Johansson (2008). Instead of regressing the log of the elapsed time to programme start within the unemployment spells of participants on a selection of the available explanatory variables that seem important for the timing of the programme, we use a post-LASSO estimator (i.e. OLS with the variables selected by LASSO estimation) to determine the relevant variables and the coefficients of this regression. We then use the estimated coefficients together with a draw from the residual distribution to predict the 'pseudo' programme starts for nonparticipants. Thus, the underlying assumption is that the assignment of programme start dates is random conditional on the variables included in the post-LASSO procedure.[20] We exclude those nonparticipants for whom this simulated start date lies outside the nine-month treatment window and those who are no longer unemployed at the assigned start date.[21] Dropping these individuals from the control group partially solves the dynamic assignment problem, because it discards individuals who tend to be not treated because they tend to leave unemployment at a high rate, and also those who are likely to be treated after the retained treatment window of nine months. Nevertheless, it does not provide a complete solution, because, in comparison to the treatment groups, the control group might still be more populated by individuals who on average exit unemployment at a higher rate *after* the (predicted) programme start, and who do not participate in the programme for this reason. To the extent that this is an issue, this would imply that we identify a lower bound for the different treatment effects (van den Berg and Vikström, 2022).

[20] The details of the determination of the pseudo programme starts can be found in Online Appendix B.

[21] One could criticize that dropping these observations can make the retained NOP population selective, and therefore bias the estimates of the GATEs and ATEs. We argue that this is not an issue if we define the population of interest to be the unemployed *at risk of programme participation.* Nevertheless, the reader must be aware that the GATEs and ATEs are biased upward if the population of interest are *all* entrants into unemployment as the dropped individuals are much more likely to be Belgian and proficient in Dutch, two features that according to our estimates reduce the effectiveness of programme participation.

### 4.4.2 Inference and multiple testing

Even if causal machine learning is a powerful technique for estimating treatment effect heterogeneity, providing correct methods of inference remains challenging. In particular, because of the multiple testing problem, it is tricky to make statements about the statistical significance of the treatment heterogeneity as reflected in the GATEs and IATEs. Semenova and Chernozhukov (2021) discuss how to construct uniform (as opposed to pointwise) confidence intervals for each IATE estimated by some debiased machine learning method, but we are not aware of results that would allow constructing uniform confidence intervals when the estimation is based on causal forests.[22]

Given the lack of a satisfactory theoretical solution to this problem, we addressed this issue in the following way. First, we only consider a limited number of GATEs that we identified ex-ante as interesting from a policy perspective. Second, the inference for the IATEs plays only a limited role in the discussions of the results. Thirdly, we discuss the results only to the extent that we consistently find the same patterns for variables of interest (like language proficiency and immigration background) across parameters (like the GATEs and k-means grouped IATEs). We are, however, aware that this is not a satisfactory solution from a theoretical perspective, and therefore mention this in the conclusion as an important avenue for future research.

[22] Crump, Hotz and Imbens (2008) provide methods in case the object of interest is a parametric or nonparametric regression model. While the parametric approach is likely to suffer from specification error, the second is subject to the curse of dimensionality in our context.

In Section 6 we calculate policy rules that aim at evaluating to what extent reassigning the unemployed to different programmes can improve the performance of the PES. To gauge their performance these calculations use the estimated IATEs. The imprecision of the estimated IATEs, therefore, translates into the measured performance gains. We take this uncertainty into account by a parametric bootstrap using the pointwise standard errors of the estimated IATEs.[23] However, the reader should be aware that the reported standard errors of the performance gains are theoretically unsatisfactory as they do not take the aforementioned multiple testing problem into account.

# 5 Results

In this section, we report the main results. We start by considering the average population effects for several outcomes of policy importance and their development over time. This informs us about the overall effectiveness of the different programmes and the dynamics of the effects. Next, we investigate whether the average population effects (ATE) differ from the effects of those unemployed workers in a particular programme (ATET). These comparisons are informative to understanding the effects of caseworkers' selection to some extent. If caseworkers select programmes that are most effective for their specific unemployed, then ATET should be larger than ATE.

Then, for the arguably most important short- and medium-run outcome, namely employment, we investigate more thoroughly the heterogeneities to the programmes and groups of unemployed by their programme participation. Subsequently, the heterogeneity of the most policy-relevant medium-run effects is investigated to a few variables considered to be of importance for the policy. Finally, in the last subsection, we present an analysis of the IATEs, i.e., the effect estimates at the finest possible level of granularity.

[23] Because of excessive computation time, we could not implement these bootstraps for the policy trees.

## 5.1 Average population effects

### 5.1.1 Dynamics and programme heterogeneity

In this section, we report the average population effects (ATE) of the different programmes in comparison with no ALMP participation (NOP) and with each other.

Figure 5.1 reports the dynamic evolution since the programme start of the effects of the different programmes in comparison with no ALMP participation (NOP) on the probability of employment.[24] Participation in short-term vocational training (SVT) modestly decreases the probability of employment only during the first four months up to 9 percentage points (pp) relative to the counterfactual of NOP. Since the average programme duration is 3.8 months, this reflects the lock-in effect. After four months the average effect on the employment probability gradually increases. It becomes positive after 11 months, and only. continues to rise to around 15 pp after 27 months, which is a substantial (statistically well determined) effect.

[24] Standard errors and confidence intervals are omitted for clarity of presentation. All effects have a standard error of about 2.3-2.5 percentage points after month 20.

*Figure 5.1: The time evolution of the ATEs of the employment probability 1 to 30 months after the programme start*

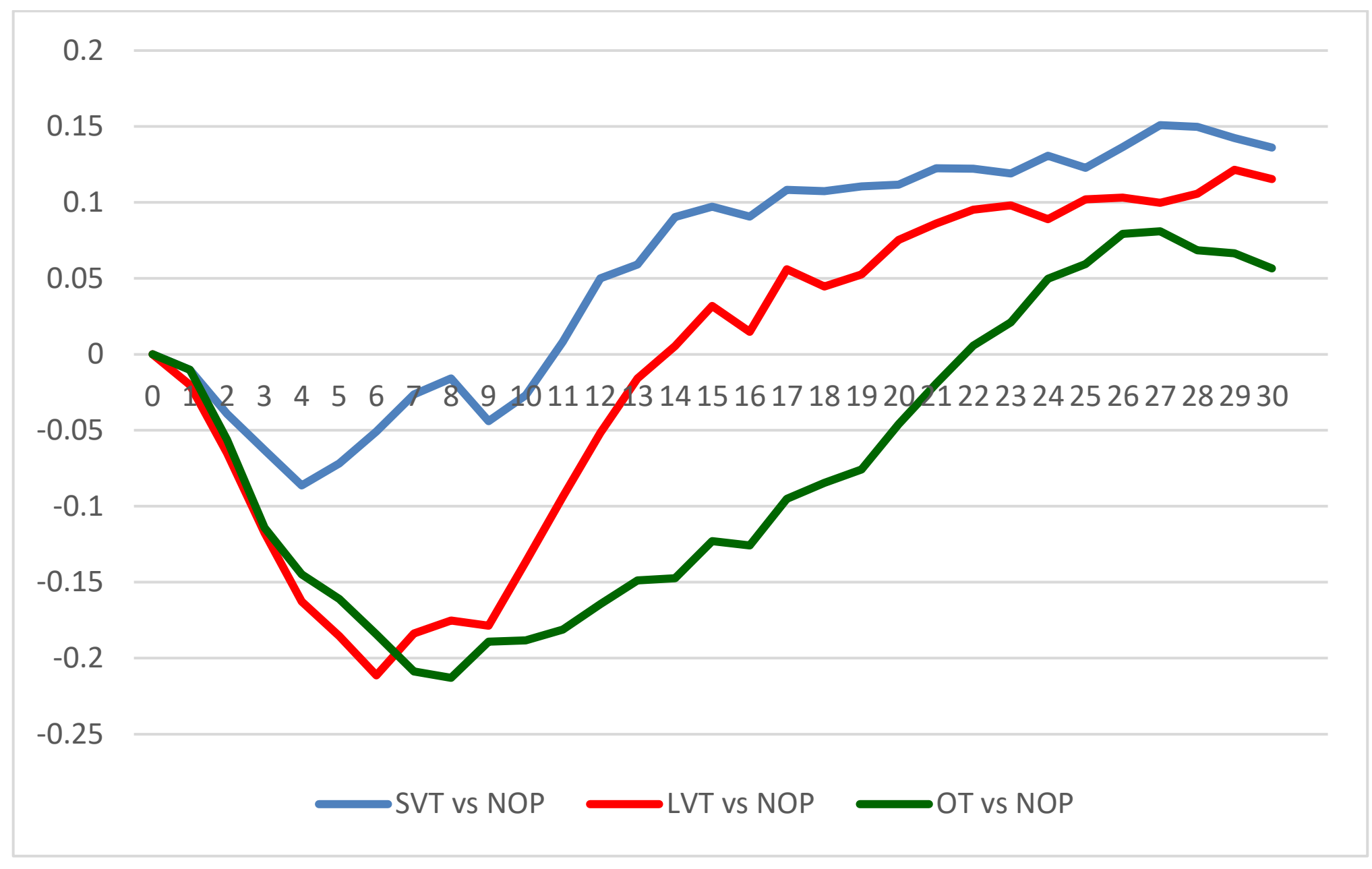

Note: The vertical axis measures the average effect on the employment probability of participating in a training programme (SVT, LVT and OT) relative to not participating in training (NOP). The horizontal axis measures the number of months after the (hypothetical) programme start.

For participants in long-term vocational training (LVT) the ATE falls much more sharply, resulting after six months in a 21 pp lower employment probability than in the counterfactual of no participation. Since the average programme duration is 7.2 months this seems to reflect again a lock-in effect. After this lock-in period, the employment probability rises rapidly for participants in LVT relative to NOP. It becomes larger than in the counterfactual of NOP after 13 months and continues to increase at a high pace until 17 months after the programme start. From then onwards the growth rate of the ATE gradually declines and it seems to stabilize at about 12 percentage points at the end of the observation period.

The impact of programme participation in LVT remains throughout the observation period below that in SVT. Unless the effect of LVT overtakes that of SVT beyond the observation period, this means that the longer time investment in human capital accumulation is not reflected in higher employment chances. It is possible that the higher time investment of LVT results in higher productivity and/or wage effects, but due to data unavailability, this could not be tested. However, Lechner, Miquel and Wunsch (2011), who also find a similar pattern for the employment effects when comparing long to short training programmes in Germany, do not find that longer training increases earnings more. Consequently, as in the German study, the finding that long training is not more effective is more likely related to the differences in the training content (see Section 3.2).[25] This explanation is also consistent with the finding in Section 6.3 that SVT is more aimed at providing an update of skills of individuals that have recent experience in a similar profession as the one at which the SVT is targeted, while LVT rather provides more fundamental re-training for a new profession. For the former type of training returns will realise faster and not necessarily less persistently than for the latter. This interpretation is also more plausible than that the effect of SVT is biased upwards due to a failure of the unconfoundedness assumption. On the one hand, such a failure is not consistent with the findings of the placebo analysis that we will report in Section 7.1. On the other hand, Altonji, Elder and Taber (2005) provide plausible theoretical arguments that, in case of a bias on unobservables, the direction of this bias should be the same as the one on observables. This would then rather, if anything, suggest a downward bias of the effects of SVT, as in Section 3.3 we presented evidence that participants in SVT are disadvantaged in terms of employability relative to those in LVT.

The negative effects during the lock-in period of orientation training (OT) are equally pronounced as those of LVT, but the lock-in effect is slightly longer, as it lasts 8 months. This long lock-in effect is presumably related to the fact that 45% of OT participants enter other programmes within 3 months after completing OT. OT (including its follow-up programmes) is less effective than the VT programmes because the corresponding ATE increases at a lower rate and attains a lower maximum – 8 pp – after 27 months. Even if one is not convinced by the absence of placebo effects reported in Section 7.1, using again the aforementioned argument of Altonji, Elder and Taber (2005), the finding in Section 3.3 that participants in OT have a comparable profile as those in LVT makes it unlikely that the smaller effects of OT would be caused by a violation of the unconfoundedness assumption. In conclusion, on average, SVT dominates LVT in terms of effectiveness, which in turn dominates OT.

[25] Note that similar to longer programmes, these shorter programmes still aim at providing skills that are useful for practising a particular profession. These are therefore distinct from job-search assistance or job-readiness programmes, and their higher short-run impact can therefore not be explained by the greater effectiveness of the latter type of programmes as documented in the meta-analysis of Card, Kluve and Weber (2018).

Next, to get a better overall picture of the effects, we investigate (i) three summary measures of the employment effects (summed up over the first and last 9 months as well as over all 30 months), and (ii) two alternative outcome measures (months in unemployment, months out-of-the-labour-force). Table 5.1 reports these estimated ATEs. The levels of potential outcomes are reported in bold on the diagonals, while the ATE of a programme in a particular row relative to the counterfactual in a column can be found off-diagonal.

*Table 5.1: Effects of the different programmes on cumulative months in employment, unemployment and out of the labour force (ATE)*

| | No ALMP participation (NOP) | Short vocational training (SVT) | Long vocational training (LVT) | Orientation training (OT) |
|---|---|---|---|---|
| Cumulative months in *employment 9 months after the programme start* | | | | |
| NOP | **3.3 (0.02)** | | | |
| SVT | -0.1 (0.1) | **3.2 (0.1)** | | |
| LVT | -1.2 (0.1) *** | -1.2 (0.2) *** | **2.1 (0.1)** | |
| OT | -1.5 (0.1) *** | -1.5 (0.2) *** | -0.3 (0.2) | **1.8 (0.1)** |
| Cumulative months in *employment between month 22 and month 30* after the programme start | | | | |
| NOP | **5.7 (0.03)** | | | |
| SVT | 1.3 (0.1) *** | **7.0 (0.1)** | | |
| LVT | 0.9 (0.2) *** | -0.3 (0.2) * | **6.6 (0.1)** | |
| OT | 0.7 (0.2) *** | -0.6 (0.2) *** | -0.3 (0.2) | **6.3 (0.2)** |
| Cumulative months in *employment 30 months* after the programme start | | | | |
| NOP | **15.9 (0.1)** | | | |
| SVT | 2.5 (0.4) *** | **18.4 (0.4)** | | |
| LVT | 0.2 (0.4) | -2.3 (0.5) *** | **0.1 (0.4)** | |
| OT | -1.4 (0.4) *** | -3.9 (0.6) *** | -1.6 (0.6) *** | **14.5 (0.4)** |
| Cumulative months in *unemployment 30 months* after the programme start | | | | |
| NOP | **11.2 (0.1)** | | | |
| SVT | -1.2 (0.3) *** | **10.0 (0.3)** | | |
| LVT | 1.4 (0.4) *** | 2.6 (0.5) *** | **12.6 (0.4)** | |
| OT | 2.9 (0.4) *** | 4.0 (0.5) *** | 1.5 (0.5) *** | **14.1 (0.4)** |
| Cumulative months out-of-the-labour force 30 months after the programme start | | | | |
| NOP | **3.2 (0.04)** | | | |
| SVT | -1.4 (0.2) *** | **1.7 (0.2)** | | |
| LVT | -1.9 (0.2) *** | -0.5 (0.3) * | **1.2 (0.2)** | |
| OT | -1.3 (0.2) *** | 0.2 (0.3) | 0.6 (0.3) ** | **1.9 (0.2)** |

Note: Outcomes are measured in months. Level of the potential outcome for the specific programme on the main diagonal in bold. All effects are population averages (ATE). Standard errors are in brackets. *, **, *** indicate the precision of the estimate by showing whether the p-value of a two-sided significance test is below 10%, 5%, and 1%, respectively.

The first panel of Table 5.1 shows that after 9 months participants in SVT accumulate on average only 0.1 months fewer months in employment than those in NOP: 3.2 months instead of 3.3 months). By contrast, participants in LVT and OT accumulate on average respectively 1.2 and 1.5 fewer months.

The second panel shows that the cumulative ATEs relative to NOP are all positive in the last 9 months of the observation window (months 22 to 30): +1.3 months for SVT, +0.9 months for LVT, and +0.7 months for OT. From the numbers reported in the other columns in this panel, we can also deduce that participants in OT and LVT accumulate less time in employment during this period than participants in SVT, with a level of significance respectively of less than 1% and 10%; the time in employment is not statistically different for participants in SVT and OT.

The third panel summarizes these employment effects over all 30 months. While participation in SVT increases the cumulative number of months in employment on average by 2.5 months relative to NOP, no significant difference is found for participation in LVT, and due to its large lock-in component participation in OT still decreases the time spent in employment by 4.4 months on average 30 months after the programme start.

The last two panels of Table 5.1 report the average total impact of the different programmes on the time spent in unemployment (UE) and out-of-the-labour-force (OLF) 30 months after the programme start. While all programmes reduce time in OLF by between 1.3 months for OT to 1.9 months for LVT, only SVT decreases the time in UE (by 1.2 months). LVT increases UE by 1.4 months, while OT increases time in UE by almost 3 months. Again, these are at least partly repercussions of the differential lock-in effects.

These findings can be rationalized with ideas in both the economic and psychological literature. First, from an economic perspective, we expect that participation in training reinforces labour force participation if the option value of participation is eventually positive, which is consistent with the findings reported in Figure 5.1. Furthermore, participating in training helps workers set clearer occupational targets because counsellors typically have such targets in mind when assigning them to training programmes. The psychological literature typically finds that goal setting leads to more time and effort spent on job search (see e.g. van Hooft and Noordzij, 2009; Latham *et al.* 2018).

### 5.1.2 Programme group heterogeneity

While in Table 5.1 we investigated the effects for the full population of unemployed, now we analyse how the effects differ for the different populations participating in the different programmes. One motivation for this perspective is that if caseworkers assign programmes according to their individual effectiveness, then we expect the effects for their own population (e.g., the effects of SVT for those participating in SVT) to be the largest. The detailed results in Table C.1 in Online Appendix C clearly show that this is not the case. Hardly any differences can be detected between the ATETs of the four treatment groups reported in the columns. This means that either the treatment effects are fairly homogeneous for these programmes or that caseworkers assign the unemployed to the different programmes without taking the individual gains of programme participation into account. Below we will show that effects are heterogeneous such that we can conclude that caseworkers fail to assign the unemployed to those programmes from which they would benefit most. In Section 6 we discuss the gains that the PES could make by improving the assignments of the unemployed to different programmes.

## 5.2 Heterogeneity of policy-relevant variables

In many situations, there are subgroups a decision-maker may particularly care about. In this section, we analyse such variables using the GATE parameter introduced above. We present the results for the overall population, as programme-population-specific effects do not appear to deviate much from the population averages. We focus on the main medium-term outcome: the cumulative number of months employed 30 months after the programme start. Of course, specifying a long list of policy-relevant variables a priori and reporting significant results bears the danger of data snooping. Here, we assume that the Flemish labour market authorities consider the following variables as particularly important: Unemployment history (last 2 and 10 years), unemployment duration at the start of the programme (below or above the median), age (younger than 25 or older than 50, below or above the median), sex, proficiency in the Dutch language (4 points Likert scale), unemployment rate in the district of residence at the start of the unemployment spell, country of birth (6 groups: Belgium, Southern EU countries, Eastern EU countries, other EU countries, Turkey or Morocco, rest of the world), and 13 education levels (from the second year of high school or below to master's degree).

Remember that we have sampled individuals who entered unemployment after they lost their job. Recent unemployment history, therefore, helps us to identify a population that is more loosely attached to the labour market. From a policy perspective, it could be interesting to identify programmes that work for such a population. A priori one could expect that the provision of vocational training may strengthen the competencies of this group and may accomplish more stable employment. By contrast, focusing on workers with little unemployment experience in the last 10 years can help identify a group of workers who had stable employment, but who lost their job abruptly. This might be the group to which OT is targeted and it is of interest to know whether such a strategy works. In Belgium, youth and older workers have difficulties in finding jobs such that it is of interest to know which policies are effective for those younger than 25 or older than 50. Discrimination both in terms of gender and migration background (country of birth and proficiency in Dutch are proxies for this) is a very sensitive political issue in Flanders and in Belgium individuals with a migration background have much more difficulty than elsewhere in the EU to find employment (Piton and Rycx, 2020). Identifying which policies work best for containing such discrimination and for getting migrants to stable employment is therefore highly relevant. In the Belgian labour market, low-educated workers are particularly at risk of unemployment such that knowledge about the relative effectiveness of policies according to the level of education of participants is valuable. Despite Flanders being a small region, unemployment rates vary substantially across districts. This is related to the limited geographic mobility within the region, induced, amongst others, by policies that heavily support home ownership and that stimulate traffic congestion. Finally, the effectiveness of training programmes according to the unemployment duration at which they start relates to the discussion of whether *preventive* or *curative* interventions are more effective.

When we test for univariate treatment effect heterogeneity of the ATEs on the cumulative number of months employed 30 months after programme start, we find statistically significant differences at the 10% level in two dimensions (proficiency in Dutch and country of birth), but not for all three programmes. Figure 5.2 illustrates the difference of the GATEs (minus ATE) of SVT relative to NOP associated with the four proficiency levels through box-and-whisker plots. The horizontal line at zero indicates the level of the ATE. One can observe a decrease in the GATEs with the proficiency level in Dutch and that the lower proficiency levels have significantly higher GATEs than the ATE. For instance, the GATE of those with no knowledge of Dutch is 1.6 months higher than the ATE (p-value below 1%).

*Figure 5.2: Difference of GATEs to ATE of SVT relative to NOP for the four proficiency levels in Dutch – Cumulative number of months employed 30 months after programme start*

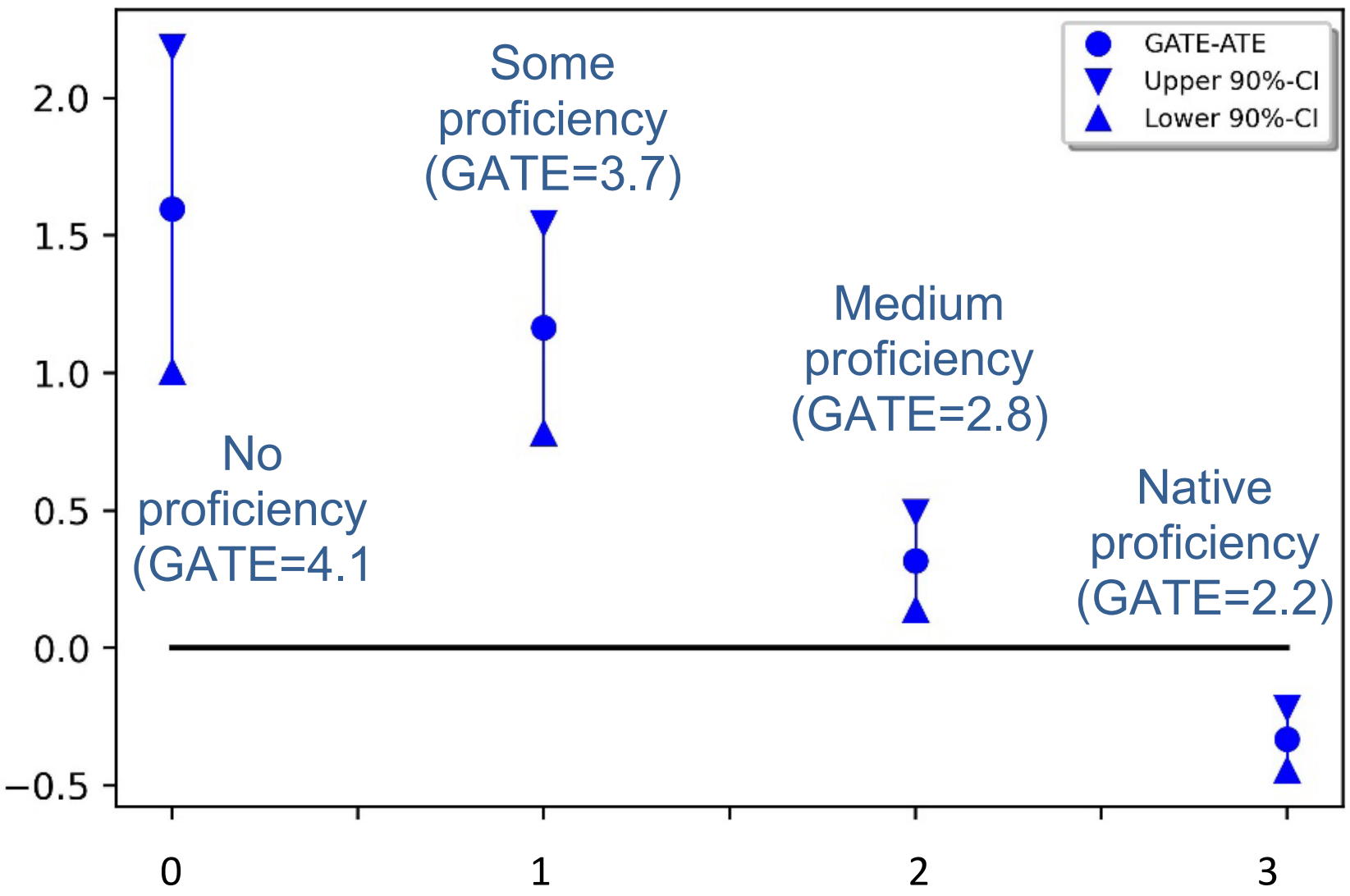

Note: Dutch proficiency is displayed on the horizontal axis. The vertical axis denotes the difference of respective GATE with ATE. (GATE-ATE) and its 90% confidence interval is shown. Dutch proficiency varies between no proficiency (0) and native proficiency (3).

The finding that the effectiveness of the programme decreases with language proficiency may seem contradictory, as a minimum requirement for training to be effective is that one can understand the instructor. However, the PES automatically provides language assistance to participants who have an insufficient understanding of Dutch.

Figure 5.3 reports the corresponding GATEs of OT versus NOP. Relative to the ATE, these GATEs display a very similar negative relationship with language proficiency as the one observed in Figure 5.2. Interestingly, while the ATE is significantly negative, the GATEs for the two lowest proficiency levels are not; only the GATEs for the two highest proficiency levels are significantly negative. The point estimates of the GATEs of LVT versus NOP also display a similar negative relationship with Dutch proficiency as reported for the other programme participations. However, none of these differences is significantly different from the ATE. We, therefore, do not display the corresponding figure.

*Figure 5.3: Difference of GATEs to ATE of OT relative to NOP for the four proficiency levels in Dutch – Cumulative number of months employed 30 months after programme start*

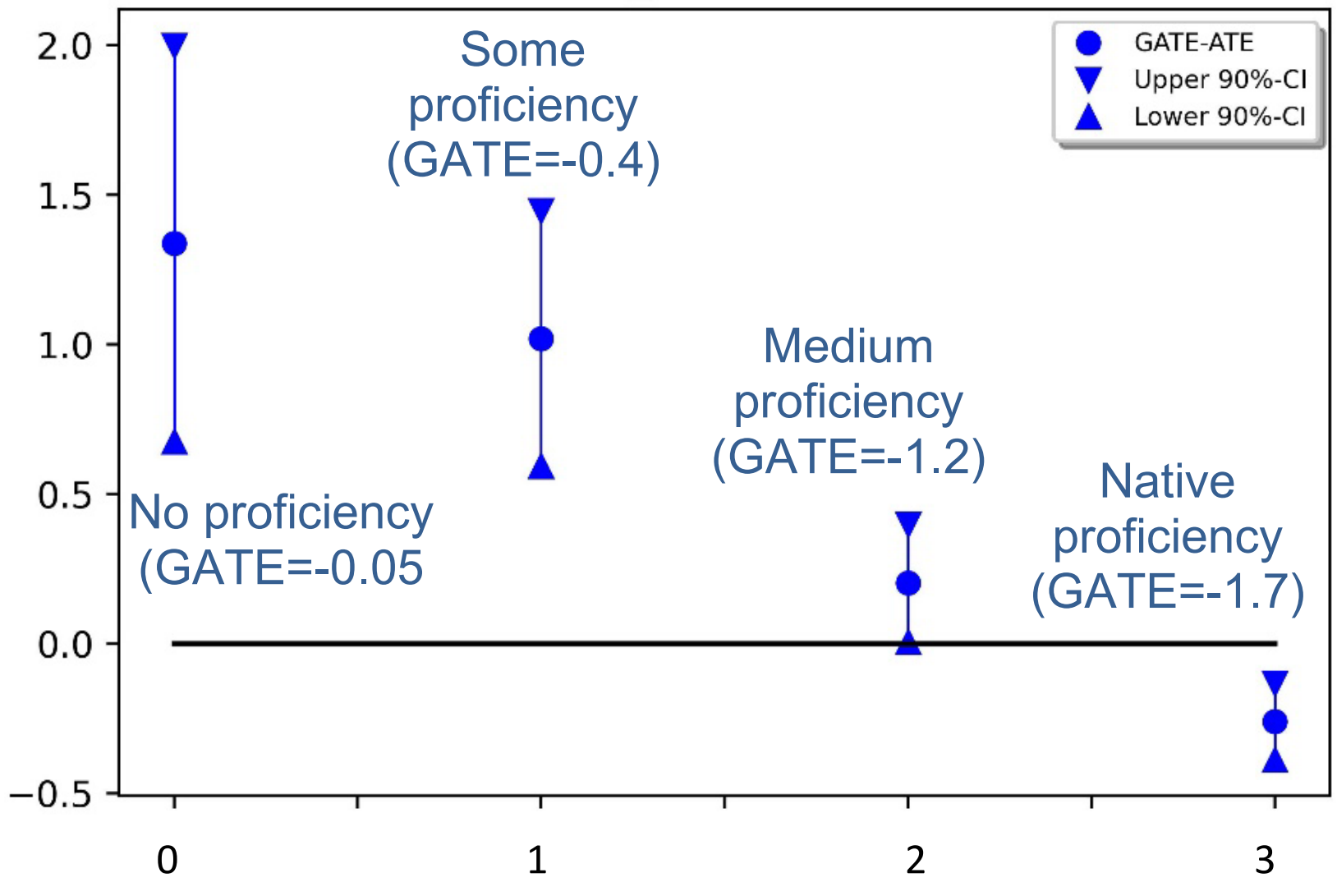

Note: Dutch proficiency is displayed on the horizontal axis. The vertical axis denotes the difference of respective GATE with ATE. (GATE-ATE) and its 90% confidence interval is shown. Dutch proficiency varies between no proficiency (0) and native proficiency (3).

Figure 5.4 illustrates how the GATEs vary by country of birth. This suggests that the GATEs of SVT relative to NOP are the highest for individuals born in Eastern European Union countries (3.6 months). It is notable that the effects for those born in Turkey and Morocco, i.e., for whom the employment rates are lower than for other foreigners, remain significantly higher (3.1 months) than for Belgians (2.1 months). Even if the precision is lower and we do not find statistically significant differences when considering the other ALMP, the patterns of the corresponding GATEs are similar and, hence, not reported. As additional evidence, we compared the GATEs for being born in Belgium or not. We find in this dimension significant differences at the 5% level for the GATEs of all three programmes relative to NOP. The GATEs for participants born in Belgium are systematically lower: 2.1 versus 3.2 months for SVT, 0.1 versus 0.5 months for LVT, and -1.7 versus -0.7 months for OT. Together with the previous finding on proficiency in Dutch this strongly suggests that SVT is more effective for migrants who recently migrated to Belgium. This conclusion also seems to hold for the other programmes but is less firmly established than for SVT.

*Figure 5.4: Difference of GATEs to ATE of SVT relative to NOP according to country of birth – Cumulative number of months employed 30 months after programme start*

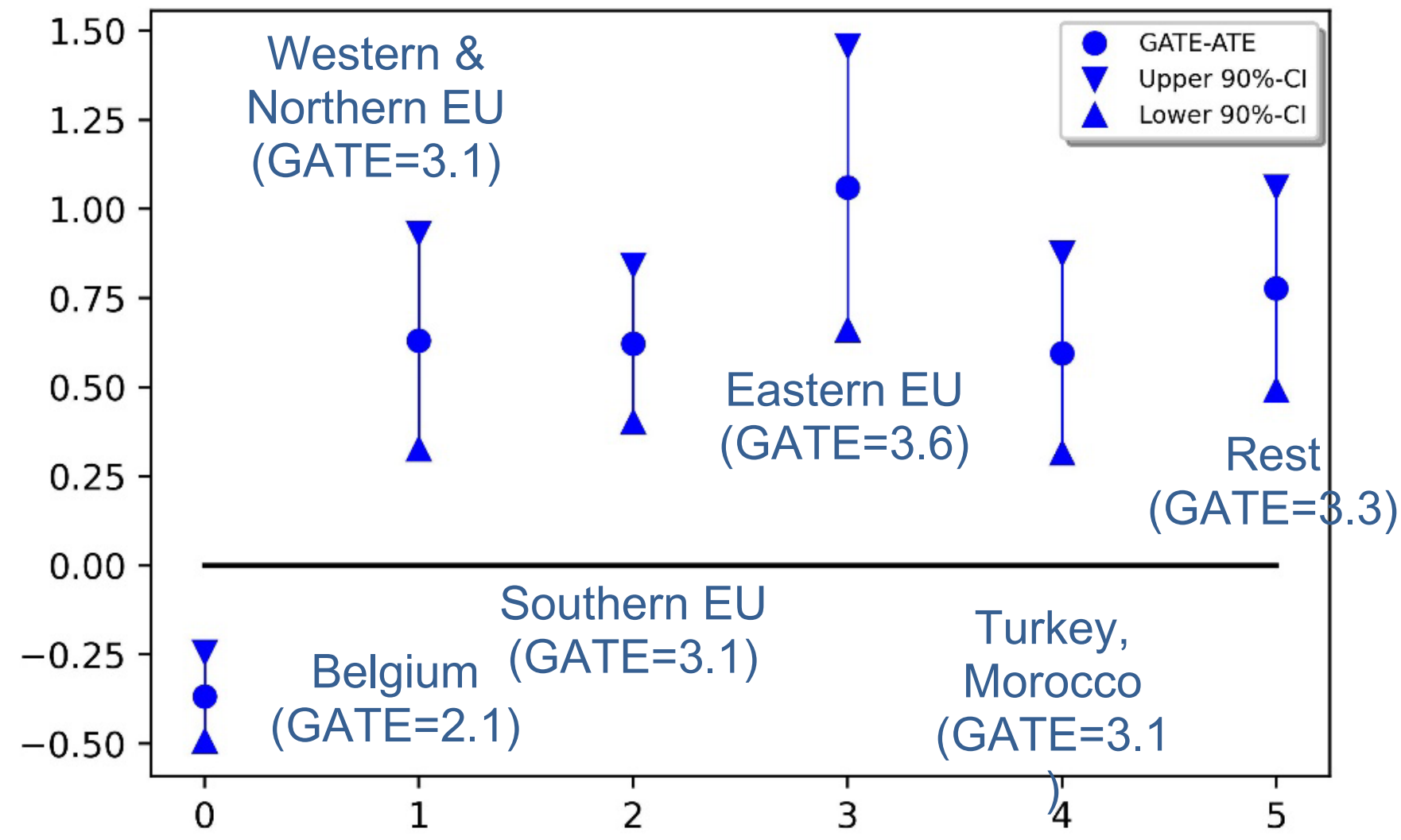

Note: Country of birth is displayed on the horizontal axis. The vertical axis denotes the difference of respective GATE with ATE. (GATE-ATE) and its 90% confidence interval is shown. The vertical axis measures the deviation of the GATE from the ATE.

## 5.3 Heterogeneity at the (averaged) individual level (IATEs)

In this section, we present the results for the individualized average effects (IATEs), which present the finest level of granularity available. To avoid flooding the reader with numbers, we will concentrate on the cumulative medium-term employment outcomes for the comparison to NOP (no programme participation) which are likely to be the most policy-relevant. We first describe the extent of heterogeneity in the programme effects. We present the results of k-means clustering analysis to get an informal characterization of sub-groups clustered according to the effectiveness of programme participation.

Figure 5.6 shows the distribution of the IATEs of SVT vs. NOP. 99.9% of the estimated effects is positive. The mean of these effects is 2.5 months (as shown in Table 5.1) and the standard deviation 1.0. About 93/88% of the estimated IATEs are significantly different from zero at the 10%/5% level. This points to two important issues: (i) There is considerable heterogeneity in the IATEs, some of which however is due to estimation error; (ii) It is much more difficult to get a precise estimate (without imposing functional forms) for the IATEs than for the GATEs and ATE that were estimated with rather high precision. These features are also visible when considering Figure 5.7, in which the sorted effects are given together with a 90%-confidence interval based on the estimated standard errors (see also Chernozhukov, Fernandez-Val, and Luo 2018). Again, we see a substantial variation of the effects, but also that the uncertainty of the ATE is much lower than for the IATEs.

*Figure 5.6: Distribution of estimated IATE of SVT vs. NOP*

Note: The horizontal axis measures the average gain of participating in SVT relative to NOP in the cumulative number of months that one is employed as measured 30 months after the (hypothetical) programme start. Kernel smooth with Epanechnikov Kernel and Silverman (normality) bandwidth.

*Figure 5.7: Overall heterogeneity: sorted effects of SVT relative to NOP – Employment 30 months after the programme start*

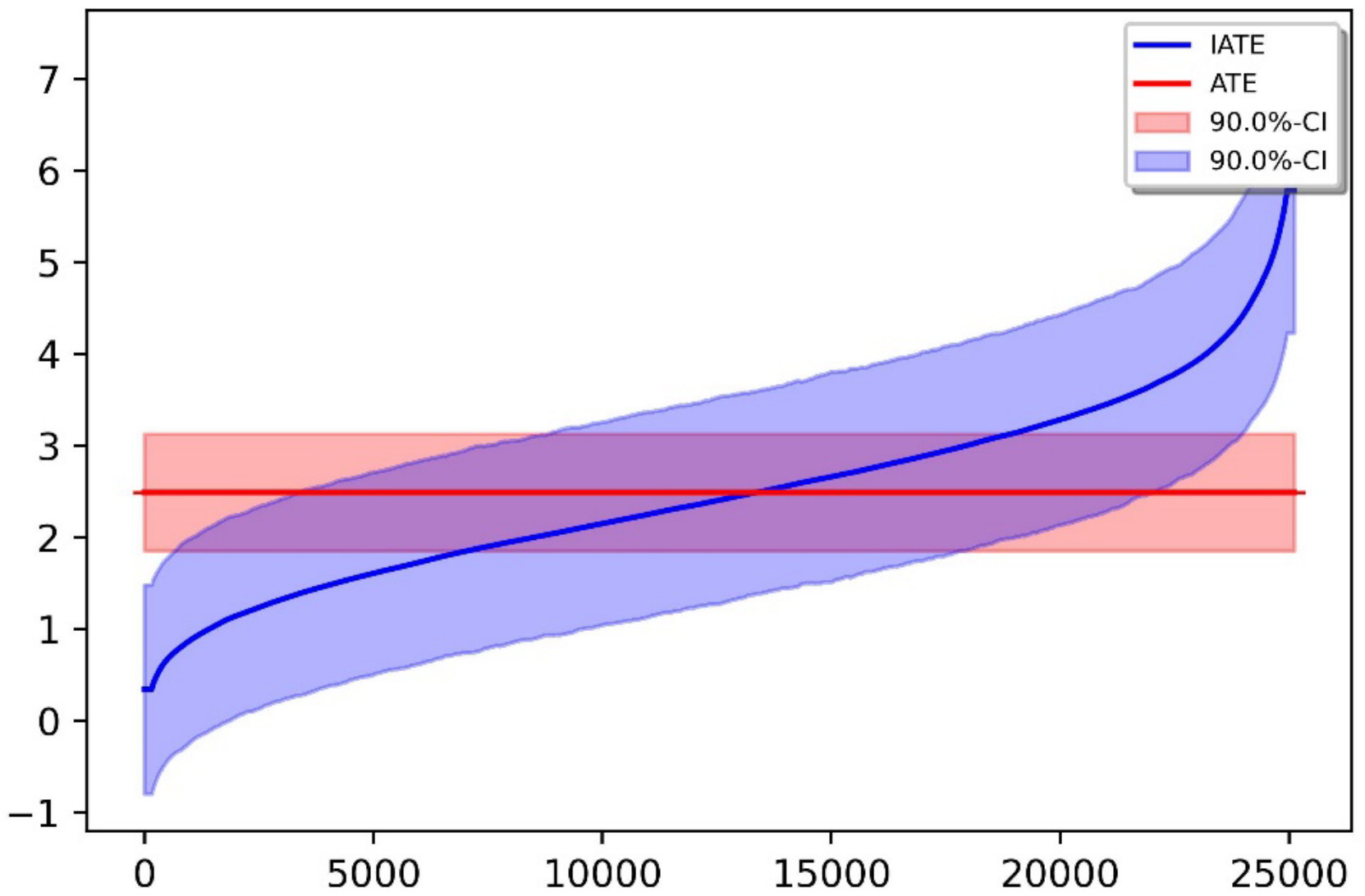

Note: IATEs are sorted according to their size. 90%-confidence interval of IATEs based on estimated standard errors and normal distribution. The vertical axis measures the average gain of participating in SVT relative to NOP in the cumulative number of months that one is employed as measured 30 months after the (hypothetical) programme start. The horizontal axis shows the rank of the ordered observations. Standard errors are smoothed by Nadaraya-Watson regression (Epanechnikov kernel with Silverman bandwidth).

The respective figures for the other programmes, as well as the sorted effects (and inference) of the difference between the IATEs to the ATE show qualitatively similar patterns to the ones presented here and are thus moved to Online Appendix C.2.

From the previous discussion of the GATEs and the distribution of the IATEs, there is substantial effect heterogeneity. In the previous section, we discussed to what extent this heterogeneity is directly related to policy-relevant variables. To be able to detect further patterns of heterogeneity at this fine level is impossible without some additional structure (due to high estimation noise otherwise). Therefore, complementing the heterogeneity analysis of the previous section, we describe the dependence of the effects on covariates by k-means++ clustering (Arthur and Vassilvitskii 2007).[26] The clustering is implemented by jointly using the IATEs of the three programme effects relative to NOP to form eight clusters. For reasons of conciseness, the clustering is only presented for the cumulative employment outcome 30 months after the programme start. The results are contained in Table 5.3.

[26] The difference between k-means++ clustering and conventional k-means clustering is an improved way to choose the initial centres of the clusters, which speeds up computation.

The clustering is close to uniformly monotone in the effectiveness of all programmes and the columns in Table 5.3 are ordered accordingly. The analysis reveals again the important heterogeneity in the programme effects. The employment gains range from 1.4 to 4.7 months for SVT, from -0.8 to +1.1 months for LVT and from -2.2 to +0.4 for OT. The programmes are the most effective for the group with the lowest proficiency in Dutch, born abroad and with very little recent unemployment *and* employment experience, on average respectively two and eight months in the last two years. This profile can only match recent entries into the labour market. The rest of the world and the Eastern EU are the most represented countries of origin in the most effective group. Taken together with the fact that the most effective clusters comprise individuals with the least recent and less recent employment and unemployment experience (and that recent school leavers are excluded from the sample), it becomes obvious that the group for which the programmes are most effective consist mainly of recent migrants. Additionally, living in a city rather than in rural areas and postponed programme starts are also associated with larger programme effects.

The two least effective groups are natives with excellent proficiency in Dutch (2.8 on average) and relatively more recent (last 2 years) and less recent (last 10 years) employment experience. Their first entry into the labour market was typically as unemployed job seekers (rather than directly in employment). It is also notable that men and younger individuals are more represented in the group for which the programmes are the least effective.

*Table 5.3: Descriptive statistics of clusters based on k-means clustering*

| Cluster | **Least beneficial** | **2** | **3** | **4** | **5** | **Most beneficial** |
|---|---|---|---|---|---|---|
| | Mean | | | | | |
| | Individualized average treatment effects (IATE) for the comparison to NOP (no participation) | | | | | |
| SVT-NOP | **1.4** | 1.7 | 2.4 | 2.7 | 3.4 | **4.7** |
| LVT-NOP | **-0.8** | 0.4 | -0.2 | 1.0 | 0.7 | **1.1** |
| OT-NOP | **-2.2** | -2.1 | -1.3 | -1.5 | -0.5 | **0.4** |
| | Selected features | | | | | |
| Age | **28** | 36 | 33 | 41 | 38 | **40** |
| Women (in %) | **33** | 52 | 39 | 56 | 51 | **55** |
| Living in a city (in %) | **24** | 25 | 38 | 34 | 47 | **47** |
| Proficiency in Dutch (3: high, 0: none) | **2.8** | 2.8 | 2.5 | 2.7 | 2.1 | **1.5** |
| Country of birth: Belgium (in %) | **96** | 92 | 66 | 69 | 34 | **18** |
| Country of birth: Western & Northern EU (in %) | **1** | 2 | 4 | 10 | 12 | **15** |
| Country of birth: Southern EU (in %) | **0** | 1 | 1 | 1 | 2 | **3** |
| Country of birth: Eastern EU (in %) | **0** | 1 | 5 | 3 | 10 | **18** |
| Country of birth: Turkey & Morocco (in %) | **0** | 1 | 5 | 5 | 11 | **10** |
| Country of birth: Rest of the World (in %) | **2** | 3 | 18 | 12 | 31 | **35** |
| BIT (unemployed at first labour market entry; in %) | **74** | 38 | 43 | 15 | 13 | **2** |
| # of months unemployed in the last *10 years* before UI | **18** | 12 | 27 | 15 | 25 | **12** |
| # of months employed in the last *10 years* before UI | **45** | 80 | 42 | 76 | 41 | **18** |
| # of months unemployed in the last *2 years* before UI | **5** | 2 | 6 | 2 | 4 | **2** |
| # of months employed in the last *2 years* before UI | **17** | 21 | 15 | 20 | 15 | **8** |
| # of days until the programme start | **63** | 77 | 92 | 105 | 124 | **165** |
| Predicted outcome without programme participation (NOP) | **16.5** | 16.3 | 15.8 | 15.7 | 15.3 | **14.9** |

Note: The outcome variable is cumulative employment in the 30 months after the programme start. All IATEs for all comparisons to nonparticipation are used to form the 8 clusters. Covariates are not used to form clusters. K-means ++ algorithm used (Arthur and Vassilvitskii, 2007).

Finally, one way to characterise the employability of the unemployed is to consider their estimated months of employment without the programme (NOP). The last row in Table 5.3 shows that in those 8 groups, programme effectiveness decreases monotonically with employability. Although this decline is moderate, this is consistent with the picture of heterogeneity revealed so far.

# 6 Policy simulations

So far, we have documented considerable heterogeneity in the effectiveness of the programmes. Do caseworkers exploit this information in the sense of assigning the unemployed to the programmes that work best for them, and, if not, to what extent could a different assignment improve the performance of the PES? To do so, we first define a criterion based on which we measure the performance of these different hypothetical allocations. We assume throughout that the policymaker pursues the following two objectives with equal weight: Maximize the number of months in employment and minimize the number of months in unemployment within the observation window of 30 months since the programme start. We then shed some light on the answers to these questions by comparing the performance of simulated hypothetical programme allocations to that of the observed allocation.

For generating new assignments, we use three different approaches:

1. Completely *random* assignment of the unemployed to the different programmes, respecting the observed capacity in terms of training slots.
2. A "*black-box*" approach that uses the estimated IATEs to allocate individuals to the programme that leads to the highest performance both with and without capacity constraints (as upper limit for the different programmes) and considering various priority rules in the latter case.
3. A *policy tree* approach that respects the capacity constraints (as upper limit for the different programmes).

The comparison to a random assignment aims at verifying the extent to which caseworkers use the information on programme effectiveness in their assignment decision. If they do, the observed allocation should perform better than the random allocation. The "black-box" approach assigns individuals to the programme that according to the point estimates perform best. We consider two main scenarios within this approach. First, we assume that the existing capacity of programmes does not impose a constraint on enrolment. This provides us with the maximum performance that can be attained by reallocating individuals to their best alternative.[27] In the second scenario, we maximize performance subject to not exceeding the existing enrolment capacity for each of the programmes. In case of excess demand, we use different priority rules to respect the capacity constraints. Individuals are not assigned to any programme in case this assignment would reduce performance relative to non-participation. Imposing these capacity constraints allows us to check whether the performance can be improved by a simple reallocation without changing the existing programme offer.

[27] Unfortunately, we do not have information about programme costs, such that we cannot conduct a fully-fledged cost-benefit analysis of such reallocation.

A problem with the "black-box" approach is that caseworkers may not trust and thus not follow such rules (when made available by an AI system), as they cannot be expected to understand how they came about (Lechner and Smith, 2007). Recently, Zhou, Athey, and Wager (2022) fill this gap by proposing a method – valid in the context of multiple treatments – that allows the derivation of policy rules based on decision trees. When restricting this method to shallow trees with a small number of nodes, the resulting rule will be simple and easy to understand and implement, or, alternatively, reveal ethical or societal concerns with the rule (e.g., Whittlestone, Nyrup, Alexandrova, Dihal, and Cave, 2019; Rambachan, Kleinberg, Mullainathan, and Ludwig, 2021). Thus, caseworkers will be more likely to follow it, at least if they do not strongly disagree. However, a potential drawback is that a shallow tree might not approximate the optimal policy as well as a black-box AI recommender system.

In Table 6.1 we report the performance gains of several simulated hypothetical allocations using our estimation results. The first two sets of columns of Table 6.1 display the description of the simulated allocation rule and the shares of the population allocated to the different programmes (in %). The third and fourth sets of columns report the performance gains of the allocation rule mentioned in the corresponding line for the population of analysis and the subpopulation of switchers. Switchers are individuals who change their treatment status under the considered hypothetical rule. Performance gains are measured according to the aforementioned criterion in deviation from the observed performance.

In calculating these policy rules, we should avoid computing them on the same data used for the estimation of the IATEs because this could, by overfitting, result in spurious performance gains. We address this concern differently for the black-box rules than for the policy trees. For the black-box rules, we re-estimate the IATEs by five-fold cross-validation using the full data set. We then apply the rules using the predictions of the IATEs in the five hold-out samples and compare the performance of these rules to the observed allocation. This procedure does not work for the policy trees because the issue of overfitting also prevents us from using the same data for building the policy trees and for evaluating their performance. We, therefore, use only a random 80% subsample (E) to estimate the IATEs, again by five-fold cross-validation. The predictions of the IATEs in the hold-out samples are then used for building the policy tree in E. We also use the estimates obtained in each iteration of the cross-validation to estimate the IATEs in the 20% random subsample (T) that was not yet used and retain the average over these iterations as the predicted IATEs in this subsample. The rule implied by the policy tree built in the E-sample is then used to assign individuals in the T-sample to programmes, and the relative performance of this assignment is then compared to the observed allocation using the aforementioned average predicted IATEs.

We report standard errors based on bootstrapping to evaluate to which extent the gains from reallocation are true, and not coming from imprecisely estimated IATEs.[28] As can be seen in Table 6.1 these standard errors are small. This enhances confidence in the reported findings.

We now compare the performance of assignment rules obtained within each of the three approaches.

---

[28] Recall, however, that we discussed in Section 4.4.2 some theoretical limitations of this bootstrap, and that excessive computation time prevented us to implement these bootstraps for the policy trees.

## 6.1 Comparing observed to random assignment

The second line of Table 6.1 reveals that random assignment performs equally well as the observed allocation. This suggests that caseworkers are unlikely to base their assignment decisions on the effect heterogeneities described in this paper. This appears to be in line with official policy: The PES of Flanders used job placement rates, like “70% of the participants must be employed within 6 months after the end of the training", as targets for evaluating caseworkers instead of targets that are based on effect sizes. Thus, the Flemish PES can improve the performance of the training programmes by adjusting the assignment according to the expected programme performance of individuals as estimated by their IATE.

## 6.2 Black-box assignment rules

Next, we consider several black-box assignment schemes that depend on the available programme capacity and the degree of certainty about the effectiveness of programmes and some priority rules. The third line (*no constraint*) shows the results for the case without any constraints. In this case, the PES would allocate nearly 99% of the unemployed to SVT, about 0.4% to LVT and no one to OT; less than 1% would remain untrained. Such an assignment could increase our performance measure by 1.73 months. This reallocation would be very costly because it would massively increase the number of SVT participants, and we lack the data to conduct a cost-benefit analysis.

The previous allocation ignores the fact that some estimated IATEs are positive (or negative) just because of estimation error. Therefore, in line four (*no constraint, only significant*) we report the outcome of a simulation in which we assign individuals only to programmes if the corresponding IATEs are significantly positive or negative at the 2.5% level of a one-sided statistical test in terms of their effect on the time spent in, respectively, employment or UE. From this simulation we see that this decreases the population that is assigned to any programme by more than 20 percentage points, but almost 76% would remain allocated to SVT, meaning that the vast majority of the IATEs of SVT are significantly positive, in line with the evidence reported in Figure 5.7. Relative to the previous scenario the performance decreases by about 10% to 1.57 months.

The next scenarios consider cases in which capacity constraints are imposed. It is assumed that the training capacity of the PES is constrained to the observed capacity (and thus programme costs remain approximately the same or will be lower if the constraint is not binding in a specific allocation).[29] Since only 6% of the unemployed participate in training, this restriction dramatically reduces the overall gains that a relocation can generate. However, for those affected by the reallocations, the would-be "switchers", the gains may be very substantial. Three scenarios are considered.

In the first scenario, priority is given to individuals with the highest returns to programme participation (*constrained, preference for largest gains*).[30] The average performance gain for the population is 4.0 days (i.e., 0.13 months). This impact is small as it is diluted by the constraint that only a maximum of 6% of the population may participate in these programmes. However, for the switchers this effect increases to 39 days on average (i.e., 1.29 months). Since this scenario results in the highest gain for all constrained scenario's, we measure the performance of all subsequent scenario's relative to this benchmark for switchers.

Interestingly, in this first constrained scenario, only 0.2% of the population is assigned to OT, while the available training slots amount to 1.9% of the population. The reason is that, for most of the population, participation in this programme harms the performance relative to the counterfactual of NOP. This problem also applies to all the next scenarios. This unused capacity would better be used to finance training slots for the other training programmes, but without information on training costs, this cannot be achieved. This means that we underestimate the gains from reallocation.

[29] Unfortunately, the average costs of the programmes are not available in the data. Therefore, we use programme shares to approximate a budget constraint.

[30] In case of excess demand for a programme, priority is given to those individuals for whom the difference in performance between the best and the second-best programme is largest. Such a rule performs better than assigning individuals to their best choice because this avoids the gain of doing so is being destroyed by the loss that a second choice for rationed individuals imposes. It might be possible to improve on this rule, but this would require taking into account the complete ordering of all treatments, which would make such a rule much more complicated.

In the second constrained black-box approach we use a priority rule that is based on a first come first served basis: In case of excess demand, individuals are randomly ordered and then assigned in this order to the best programme in which there is still capacity available until all available programme slots are filled. Such a rule could be justified based on an equality of opportunity criterion. However, this rule attains only 53% (0.68/1.29 = 0.53) of the gains that switchers can get relative first to the previous benchmark scenario.

Finally, we consider a scenario in which the priority to be assigned to the best available programme is given to those individuals with the highest number of months in unemployment over the last 10 years before entry into unemployment, a rule that could be justified by concerns for equity. In this case, the performance increases somewhat to 56% of that of the benchmark scenario.

*Table 6.1: Overall performance gain of simulated hypothetical programme allocations (equal weight for an increase in months in employment and reduction in months in unemployment)*

| | Share of different programmes in % | | | Performance gain for all | | Performance gain for switchers | |
|---|---|---|---|---|---|---|---|
| | SVT | LVT | OT | Mean | SE | Mean | SE |
| Observed | 2.3 | 2.0 | 1.9 | - | - | - | - |
| | | | | Deviation from the observed state | | | |
| Random (restricted) | 2.3 | 2.0 | 1.9 | 0.012 | 0.002 | 0.10 | 0.018 |
| Black-box – no constraint | 98.7 | 0.4 | 0.0 | 1.73 | 0.004 | 1.78 | 0.004 |
| Black-box – no constraint, only significant | 76.2 | 0.3 | 0.0 | 1.57 | 0.005 | 2.05 | 0.007 |
| Black-box – constrained, preference to largest gains*) | 2.3 | 2.0 | 0.2 | 0.13 | 0.002 | 1.29 | 0.020 |
| Black-box – constrained, first come, first served**) | 2.3 | 2.0 | 0.5 | 0.07 | 0.002 | 0.68 | 0.020 |
| Black-box – constrained, preference to lots of past UE***) | 2.3 | 2.0 | 0.5 | 0.08 | 0.002 | 0.72 | 0.019 |
| Policy tree 2 levels, constrained | 2.4 | 0.0 | 0.0 | 0.11 | - | 1.31 | - |
| Policy tree 3 levels, constrained | 2.4 | 2.2 | 0.0 | 0.12 | - | 1.11 | - |
| Policy tree 4 levels, constrained | 2.0 | 2.3 | 0.0 | 0.11 | - | 1.07 | - |

Notes: *Performance gain* measures the equally weighted sum of the average number of the gained months in employment and the average number of lost months in unemployment relative to the NOP or observed state within 30 months of the programme start. *Switchers* are individuals who change their treatment status under the considered rule; *) In case of excess demand for a programme, priority is given to those individuals for whom the difference in performance between the best and the second-best programme is the largest. **) In case of excess demand, individuals are randomly ordered and then assigned in this order to the best programme in which there is still capacity available until all available programme slots are filled; ***) In case of excess demand, priority is given to the highest number of months in unemployment over the last 10 years before entry in unemployment. SE = standard errors based on bootstrapping.

## 6.3 Assignment rules based on shallow decision trees

Next, we take up the proposal by Zhou, Athey, and Wager (2022) to use shallow decision trees to obtain more intuitive, interpretable rules. We consider trees of depth 2 (resulting in 4 strata with potentially different allocations), of depth 3 (8 strata) and 4 (16 strata). The allocations are obtained by a slight modification of Algorithm 2 of Zhou, Athey and Wager (2022) which should lead to some better allocations at the expense of somewhat higher computational costs. The original paper does not discuss how to deal with categorical variables or with constraints which are both important issues in this setting. The details of these aspects, and how the constraints are enforced, are documented and explained in Online Appendix B.4. The case without constraints is not discussed, because it assigns virtually all unemployed to SVT and is therefore not interesting.

As can be observed in Table 6.1, the capacity constraints, which should be seen as an upper bound on the share of participants in the different groups, are not exactly satisfied in the test sample. This is the consequence of two issues. First, it may be optimal not to use a programme because its effects are not large enough. Second, since the constraints are only enforced during tree building, they may be violated in the test sample. This explains why in the policy trees of depths 3 and 4 the fractions do not exactly match the capacity constraints for SVT and LVT. As the main objective of this paper is not to find optimal procedures for building policy trees, we leave it for further research to potentially refine the procedures to deal with the mentioned issues.

Even if these deviation from the exact capacity constraints do not allow us to make precise comparisons between the black-box rules and the policy trees, it is clear from the table that a simple rule based on a tree of depth 2 performs already well compared to the constrained black-box approach with a preference for largest gains, i.e., the aforementioned benchmark scenario. Even if this rule assigns nobody to SVT and OT, and only marginally more than the available capacity for SVT (2.4% instead of 2.3% of the population), this simple rule attains about 85% $(0.11/0.13 = 0.85)$ of the benchmark gains. Policy trees of 3 or 4 levels only marginally improve on this simple policy tree.[31]

Table 6.2 reports the associated assignment rules for the shallow decision trees. In the policy tree of depth 2 only unemployed with less than 6 months of employment experience in Belgium in the last 10 years and born in Eastern EU, Turkey, or Morocco are assigned to SVT. Nobody else is assigned to a programme. This rule is very much in line with our findings in Section 5.2, where we found that the GATE of SVT was very high for a group which we identified as *recent migrants*. The fact that we select individuals with less than 6 months of employment experience in the last 10 years implies that the migration must have occurred only recently for the selected unemployed and that these workers should have accumulated some recent employment experience abroad because eligibility for unemployment benefits requires proof of an employment spell of at least one year, possibly abroad (see Section 2). This is also consistent with the finding reported in Figure 5.2 that the effectiveness of SVT decreases monotonically with the level of proficiency in Dutch: recent migrants are unlikely to be proficient in Dutch. Moreover, in Figure 5.4 we reported that the GATE of SVT was the highest for those who were born in Eastern Europe.

[31] The much higher performance gains of the switchers for the 2-level policy tree can be explained by the fact that the rule only assigns individuals to the best performing programme (SVT) and nobody to the other programmes. In the higher-order policy trees switchers are also assigned to the lower-performing SVT. We cannot, therefore, compare the performance of the switchers between the different policy trees.

For the policy trees of depths 3 and 4, it appears that younger workers within the group of recent migrants[32] are more likely selected in SVT, while older workers *not* restricted to the group of recent migrants are selected in LVT: the age thresholds are 40 and 34 years in the trees of depth 3 and 4, respectively. Those selected in LVT are not only older, but they are also typically workers with little recent employment experience who have been employed in specific sectors. This allocation rule suggests that LVT is most effective for older workers who lost substantial human capital after being displaced from a job in specific sectors, and who did not find employment in the recent years before entry into unemployment. This makes sense because such workers are most likely to be affected by a structural external shock that requires professional reorientation, and, hence, long vocational training. The finding that this long vocational training is tailored to this target group with little recent employment experience and SVT more to a group with more recent employment experience abroad may also explain why the performance of SVT dominates that of LVT (see Figure 5.1): The LVT may be more designed to fundamentally retrain individuals for a different profession, while SVT may be more aimed at refreshing and updating skills of workers who may already have already acquired similar skills but in a different context, such as in a same profession abroad.

[32] Recent migrants are identified somewhat differently in the policy rules of depth 3 and 4 than in the 2-level tree: including those born in the 'rest of the word' instead of those born in Turkey or Morocco in the tree of depth 3, and those not born in Belgium and no proficiency in Dutch in the tree of depth 4. Also, the identification of little past employment experience is identified in slightly different ways across these trees.

*Table 6.2: Assignment rules of shallow decision trees*

| Tree depth = 2 | | | |
|---|---|---|---|
| Short training (SVT) | Long training (LVT) | Orientation training (OT) | No programme participation (NOP) |
| • Worked ≤ 6 months in last 10 years<br>• Born in Eastern EU or Turkey or Morocco | • Nobody | • Nobody | All others |
| Tree depth = 3 | | | |
| • Worked ≤ 20 months in last 2 years<br>• Age less than 40 years<br>• Born in Eastern EU, or 'Rest' | • Worked ≤ 20 months in last 2 years<br>• Age at least 40 years<br>• Last employment in specific sectors | • Nobody | All others |
| Tree depth = 4 | | | |
| • Age less than 34 years<br>• Did not work in last 2 years<br>• Not born in Belgium<br>• No knowledge of Dutch | • Age at least 34 years<br>• Did not work in last 2 years<br>• Worked less than 104 months in last 10 years<br>• Last employment in specific sectors | • Nobody | All others |

Note: *For the sake of brevity, we do not list the specific sectors selected by the decision tree.*

To conclude, shallow decision trees seem promising as they provide simple and transparent rules that can attain a performance level of about 85% of the best corresponding black-box rule. Nevertheless, the use of such trees may still be contested on different grounds. For instance, the use of country of birth and knowledge of Dutch may not be legally or politically acceptable. However, even if this is an issue, this does not compromise the optimal policy approach. Quite on the contrary, this approach makes the rule more transparent, which is key in making its use socially acceptable or in defining restrictions that could be fed into the algorithm to determine a simple rule that is ethically and politically acceptable (see e.g., Devin and Sydnor, 2011).

# 7 Sensitivity analysis

## 7.1 Placebo analysis

To further convince the reader (and us) that the matching variables sustain the CIA, we provide two placebo validation tests. The first one is like the one proposed by Imbens and Wooldridge (2009, pp. 48–50). This validation consists in estimating with the same methodology the ATEs of participating in a future training programme within a preceding unemployment spell. Since (unanticipated) future participation in training should not have any impact on the current outcomes, finding an effect close to zero provides some support for the CIA. Because the first placebo test is based on less than a third of the sample used in the main analysis, is highly selective in terms of prior labour market history, and the evaluation period of 9 months is rather short, we also implement a second placebo test.[33] This second test has the advantage of being much less selective (77% of the sample is retained), and of making the placebo evaluation period as long as in the main analysis, i.e., 30 months.

To implement the first placebo test, we select from the population of analysis the subpopulation that has experienced at least one unemployment spell – in case of multiple spells, we retain the first observed one – starting between September 2008 and February 2014. February 2014 is used, because it leads to a gap of 9 months between the start of the last unemployment spell that was retained for the placebo sample and the first considered entry in the main analysis, i.e., December 2014. This gap allows us to estimate the placebo treatment effects during the 9 months since the start of the preceding unemployment spell. This choice of

[33] We thank an anonymous reviewer for suggesting this test.

9 months aims at finding a balance between not reducing the size of the placebo population too much – this size declines rapidly with the size of the gap time – and having a sufficiently long period over which to measure the placebo effects. To avoid contamination, we dropped all individuals who entered an ALMP during this preceding unemployment spell. The eventual sample on which this placebo analysis is conducted consists of 17,943 non-participants, and 360, 336 and 285 participants in SVT, LVT and OT, respectively.

*Table 7.1: Placebo Effects for the different future programmes on cumulative months in employment, unemployment and out of the labour force (ATE)*

| | No ALMP participation (NOP) | Short vocational training (SVT) | Long vocational training (LVT) | Orientation training (OT) |
|---|---|---|---|---|
| Cumulative months in employment 9 months after entry in the preceding unemployment spell | | | | |
| NOP | **3.9 (0.1)** | | | |
| SVT | 0.01 (0.3) | **3.9 (0.3)** | | |
| LVT | 0.5 (0.3) | 0.5 (0.4) | **4.3 (0.3)** | |
| OT | 0.001 (0.4) | -0.02 (0.4) | 0.5 (0.4) | **3.9 (0.2)** |
| Cumulative months in unemployment 9 months after entry in the preceding unemployment spell | | | | |
| NOP | **4.8 (0.04)** | | | |
| SVT | -0.1 (0.3) | **4.8 (0.3)** | | |
| LVT | -0.4 (0.3) | -0.5 (0.4) | **4.4 (0.3)** | |
| OT | -0.002 (0.3) | -0.1 (0.4) | 0.4 (0.4) | **4.8 (0.3)** |
| Cumulative months out of the labour force 9 months after entry in the preceding unemployment spell | | | | |
| NOP | **0.4 (0.01)** | | | |
| SVT | -0.1 (0.1) | **0.3 (0.02)** | | |
| LVT | -0.1 (0.1) | -0.005 (0.1) | **0.3 (0.1)** | |
| OT | -0.03 (0.1) | 0.04 (0.2) | 0.04 (0.2) | **0.3 (0.1)** |

Note: Outcomes are measured in months. The level of the potential outcome for the particular programme is on the main diagonal in bold. All effects are population averages for the respective placebo programme participants given in the column. Standard errors are in brackets. *, **, *** indicate the precision of the estimate by showing whether the p-value of a two-sided significance test is below 10%, 5%, and 1% respectively.

Table 7.1 reports the results for three outcomes: cumulative number of months employed, unemployed and out-of-the-labour force 9 months after entry in the preceding unemployment spell. The results show that all ATEs are close to zero and precisely estimated despite the rather small programme groups. Moreover, we do not find evidence of effect heterogeneity in any dimension, in particular not in the country of birth, Dutch proficiency, or employment experience within the last two years, which are dimensions that matter most in the main analysis.

In the second placebo test, we retain all individuals in the sample for whom at least 30 months of labour market history before the start of the unemployment spell is available. We then estimate the ATEs on this subsample during the 30 months preceding entry in the same way as for the 30 months since the programme start. These ATEs are all very close to zero, providing again support for the CIA. More details can be found in Online Appendix E.

## 7.2 Tuning parameters

To investigate the stability of the MCF estimates to various tuning parameters (see also Online Appendix B for more details), we performed the following sensitivity exercises: (i) The number of bootstrap replications has been increased from 1000 to 2000 replications; (ii) the minimum leaf size has been varied from 5 to 3 and 7; (iii) the subsampling share was decreased from 67% to 50%; (iv) the number of variables used for splitting any particular leaf has been varied; (v) estimation was performed with and without prior deselection of irrelevant features, and (vi) the penalty term in the MCF objective function has been increased 10 fold from its base value that equals the variance of the respective outcome variable. None of these variations led to any substantial changes in the estimation results. Note also that the estimations in the previous version of the paper were based on Gauss code (Aptech Inc.) and led to the qualitatively same result which may be seen as a safeguard against programming errors.

## 7.3 Comparison to propensity score matching

To check that our approach does not deliver nonsense results, we compared the MCF findings for the ATE with conventional matching estimates based on propensity-score-matching-with-bias-adjustment, an estimator suggested by Lechner, Miquel, and Wunsch (2011) that was demonstrated to perform well in the extensive Empirical Monte Carlo study of Huber, Lechner, and Wunsch (2013). While the detailed results are shown in Online Appendix C.4, overall point estimates, as well as standard errors, lead to very similar conclusions as for the MCF. Of course, such a comparison could only be done for the ATE (or ATEs for some large subgroups), as such estimators are not adapted for the estimation of more disaggregated effects.

# 8 Conclusion

In this paper we used causal machine learning to investigate the average and heterogeneous effects of recent training programmes in Belgium, using administrative individual data from the Public Employment Service of Flanders. We found that on average all programmes have positive employment effects in the medium run, although for orientation training (OT) these positive effects came too late and are too small to compensate for the early negative effects in the lock-in period. It turned out that on average short vocational training is more effective than longer vocational training courses as well as orientation training. Analysing the heterogeneity of the effects, the striking result appeared that programmes work better (even after the lock-in period) for unemployed with low employability, in particular, recent migrants with limited language skills. Using the fine-grained results for analysing the assignment policy of Flanders' public employment service revealed that different rules for allocating the unemployed to existing programme slots could lead to a substantial improvement in labour market performance at no or small additional costs.

We may compare these findings with the meta-analysis of Card, Kluve and Weber (2018) who analysed the effectiveness of active labour market policies (ALMP) based on more than 200 papers. In line with their general findings, our analysis detects effects that are close to zero in the short-run due to lock-in, but positive later on. They generally find that programmes with more human capital accumulation (i.e., training) are more effective in the longer run. Our findings add nuance to these conclusions as our evidence shows that SVT is as effective as LVT even in the long run. Card et al. (2018) find that heterogeneity is relevant, but they do not report, as we do, higher effectiveness for (recent) migrants. They report evidence of higher impacts for women, long-term unemployed, and during recessions. We do not find a differential impact for women. Although we cannot measure how the effects evolve over the business cycle, we do not find any evidence that residence in high unemployment regions moderates the treatment effects. We find that the treatment effect decreases monotonically with employability rather than that it increases specifically with elapsed unemployment duration at the programme start. This effect heterogeneity is found in particular during the lock-in phase because then more employable individuals would have found a job more easily in the absence of participation in training than less employable individuals.

We used our estimates of individual programme effects to evaluate the existing allocation of the Flemish PES to the considered training programmes. We found that changing the assignment rules without changing the offered training capacity could, for the group affected by this reassignment, on average increase, respectively decrease, the time spent in employment, respectively unemployment, by about one month and one week within 30 months of the programme start. This is a substantial gain, even more so because as in this reassignment most of the training capacity of the orientation training remains unused as it performs worse for most individuals than not participating in any training. Therefore, if these training slots could be redirected to one of the other training programmes, gains could be reinforced. This illustrates the social value of using CML methods in the allocation of the unemployed to active labour market policies.

Another important message is that we have shown that a shallow policy tree of depth 2 can define a simple rule that attains already about 85% of the aforementioned gain obtained by a “black-box” rule that exploits all the information available in the estimation results. One advantage of simple rules is that caseworkers are more likely to implement them because they can easily be understood. Another advantage is that the rule becomes more transparent and, hence, more ethically or societally acceptable (Whittlestone, Nyrup, Alexandrova, Dihal, and Cave, 2019; Rambachan, Kleinberg, Mullainathan, and Ludwig, 2021). However, in this application, the simple rule requires discrimination on grounds – country of birth and knowledge of Dutch – that might not be legally or politically acceptable. If so, this does not compromise the approach, but merely defines restrictions that could be fed into the algorithm that determines the simple rule.

Future work could address many open issues, such as extending the database with (i) programme costs so that the derivation of optimal policy rules can be based on the net social value of programmes, and with (ii) additional control variables, so that the effects of other programmes of the active labour market policy of Flanders that are ignored here can be credibly evaluated as well. As mentioned in the introduction, the gains to reallocation can only increase by including more programmes in the choice sets of the reassignment rules. Furthermore, it will be interesting to see whether similar heterogeneity appears in other countries with comparable policies. More generally, extending the CML frame to allow it to address issues related to dynamic assignment and multiple testing problems is likely to lead to additional insights into policy effectiveness and assignment optimality.